\documentclass[
]{ceurart}

\usepackage{listings}
\usepackage{multirow}
\usepackage{enumitem}
\usepackage{caption}
\usepackage{subcaption}
\usepackage{stfloats}
\usepackage{booktabs}  
\usepackage{graphicx}  
\usepackage{array}     

\begin{document}

\copyrightyear{2024}
\copyrightclause{Copyright for this paper by its authors.
  Use permitted under Creative Commons License Attribution 4.0
  International (CC BY 4.0).}

\conference{IntRS'24: Joint Workshop on Interfaces and Human Decision Making for Recommender Systems, October 18, 2024, Bari (Italy)}

\title{Designing and Evaluating an Educational Recommender System with Different Levels of User Control}


\author[]{Qurat Ul Ain}[%
email=qurat.ain@stud.uni-due.de,
]
\cormark[1]

\author[]{Mohamed Amine Chatti}[%
email=mohamed.chatti@uni-due.de,
]
\cormark[1]

\author[]{William Kana Tsoplefack}[%
email=william.kana-tsoplefack@stud.uni-due.de,
]

\author[]{Rawaa Alatrash}[%
email=rawaa.alatrash@stud.uni-due.de,
]

\author[]{Shoeb Joarder}[%
email=shoeb.joarder@uni-due.de,
]

\address[]{Social Computing Group, Faculty of Computer Science, University of Duisburg-Essen, Duisburg, Germany}

\begin{abstract}
Educational recommender systems (ERSs) play a crucial role in personalizing learning experiences and enhancing educational outcomes by providing recommendations of personalized resources and activities to learners, tailored to their individual learning needs. However, their effectiveness is often diminished by insufficient user control and limited transparency. To address these challenges, in this paper, we present the systematic design and evaluation of an interactive ERS, in which we introduce different levels of user control. Concretely, we introduce user control around the input (i.e., user profile), process (i.e., recommendation algorithm), and output (i.e., recommendations) of the ERS. To evaluate our system, we conducted an online user study (N=30) to explore the impact of user control on users' perceptions of the ERS in terms of several important user-centric aspects. Moreover, we investigated the effects of user control on multiple recommendation goals, namely transparency, trust, and satisfaction, as well as the interactions between these goals. 
Our results demonstrate the positive impact of user control on user perceived benefits of the ERS. Moreover, our study shows that user control strongly correlates with transparency and moderately correlates with trust and satisfaction. In terms of interaction between these goals, our results reveal that transparency moderately correlates and trust strongly correlates with
satisfaction. Whereas, transparency and trust stand out as less correlated
with each other.

\end{abstract}
\begin{keywords}
  Educational Recommender Systems \sep
  Interactive Recommender Systems \sep
  User Control \sep Transparency \sep Trust
\end{keywords}
\maketitle

\section{Introduction}
Recommender systems (RSs) are widely used across various application domains, such as e-commerce sites, online streaming websites, and social media platforms. These systems have proven effective at enhancing user experience and aiding decision-making through personalized recommendations. In recent decades, RSs have also been applied to the field of education, leading to the development of educational recommender systems (ERS) \cite{manouselis2011recommender, khanal2020systematic}. In this context, RSs are for example used to create personalized learning experiences \cite{valtolina2024design}, recommend suitable formal or informal learning materials \cite{chau2018learning}, suggest MOOCs \cite{BOUSBAHI20151813}, and adapt to context-aware learning environments \cite{santos2016toward}.

Conventional RSs usually offer minimal feedback options in the user interface, permitting users merely to indicate if they like/dislike a recommendation \cite{jin2018effects}. Interactive RSs (IntRSs) have been into the limelight as an approach to empower users to control and interact with the RS \cite{he2016interactive, jugovac2017interacting, jannach2017user}. Controllability refers to the extent to which the system allows users to adjust the recommendation process to enhance the quality of recommendations \cite{he2016interactive}. Concretely, the users can control the RS at three different levels, namely interacting with the input (i.e., user profile), process (i.e., recommendation algorithm), or output (recommendations) of the RS \cite{he2016interactive,harambam}. 
The control is provided in the RS by allowing users to adjust preferences, change parameters of the underlying algorithm, directly interact with recommendations, and provide feedback, resulting in greater perceived control and a more transparent recommendation process \cite{knijnenburg2012inspectability}. Compared to the application of IntRSs in e-commerce, entertainment, and social media domains, providing user control is under-explored in ERSs \cite{barria2019explaining}.

User control has proven to have positive impact on different recommendation goals. These include perceived accuracy of recommendations \cite{schaffer2015hypothetical, Donovan, 2012tasteweights, maxwell2015, smallworlds}, usability \cite{tsai2017providing, Kangasrasio2015, bruns2015should, zhao2010}, perceived usefulness \cite{tintarev2015inspection, Kangasrasio2015, chen2012cofeel, zhao2010}, user-perceived transparency \cite{tsai2017providing, tsai2021effects}, trust \cite{schaffer2015hypothetical, bruns2015should, loepp2014choice}, user experience \cite{knijnenburg2012inspectability, Donovan, 2012tasteweights, setfusion2014}, cognitive load and recommendation
acceptance \cite{jin2018effects}, user satisfaction \cite{tsai2017providing, jin2016go, linkedvis2013, smallworlds, talkexplorer}, and user acceptance \cite{Kangasrasio2015, linkedvis2013}.
These findings highlight the impact of user control on various recommendation goals, suggesting that these goals may interact with each other. However, to the best of our knowledge, no work has yet investigated the impact of user control on transparency (i.e, explain how the system works), trust (i.e., increase user's confidence in the system), and satisfaction (i.e., increase the ease of use or enjoyment) \cite{tintarev2015explaining} together, nor has there been an investigation into how these goals interact with each other in an interactive recommendation context.

This paper addresses these gaps by introducing user control at multiple levels within the ERS module of the MOOC platform CourseMapper \cite{Ain2022}. In this way, we enable users to interact with the input, process, and output of the ERS. Moreover, we present the systematic design of the ERS which is also lacking in the existing literature on IntRS and ERS. Furthermore, we examine the impact of user control on multiple recommendation goals, namely transparency, trust, and satisfaction. We also explore how these goals interact with each other. The following research questions guide our investigation:
\begin{itemize}
    \item RQ1. How does complementing an ERS with user control impact users' perceptions of the ERS? 
    \item RQ2. What are the effects of user control on transparency of, trust in, and satisfaction with the ERS?
    \item RQ3. How do the recommendation goals of transparency, trust, and satisfaction interact with each other in an interactive recommendation setting?
\end{itemize}

To answer these research questions, we conducted an online user study (N=30). 
Our findings demonstrate the positive impact of user control in ERSs
in terms of several important user-centric aspects including perceived accuracy, novelty, interaction adequacy,  perceived user control, transparency, trust, user satisfaction, and use intentions. Moreover, our analysis shows that user control has at least moderate correlation with all goals, while some pairs are particularly
strongly correlated with each other. More specifically, user control strongly correlates with transparency and moderately correlates with trust and satisfaction. Referring to the interaction between these goals, our study indicate that while transparency moderately correlates with satisfaction and trust strongly correlates with satisfaction, transparency and trust stand out as less correlated with each other. 

The remainder of this paper is organized as follows. We first introduce two branches of related work, namely educational recommender systems and interactive recommender systems  in Section \ref{relatedwork}. We then describe the systematic design and implementation of our interactive ERS in Section \ref{design}. Next, we present the details of the online user study that we conducted to evaluate our ERS in Section \ref{evaluation} and the analysis of the results in Section \ref{results}. 
Finally, in Section \ref{conclusion}, we summarize the work and outline our future research plans.
\section{Background and Related Work}\label{relatedwork} 
\label{background}
This section discusses related work on the application of recommender systems in the educational domain and interactive recommender systems that support user interaction with and control of recommender systems.
\subsection{Educational Recommender Systems} \label{ERS}
Educational recommender systems (ERSs) have become a vital tool in personalized learning environments, offering tailored recommendations to enhance the educational experience. These systems leverage various algorithms and data sources to suggest resources, courses, and learning activities that align with individual needs, preferences, and learning styles \cite{ain2024learner}. 
ERSs have been facilitating learning and teaching in various ways. For example, to recommend learning materials to support instructors in online programming courses \cite{chau2018learning}, to recommend educational activities to a group \cite{fotopoulou2020interactive}, and to provide recommendations while preparing for the oral examination of a language learning course \cite{santos2016toward}. Most of these ERSs either propose algorithmic enhancements or new frameworks for recommendation, or implement existing/new recommendation techniques in an educational context,
such as collaborative filtering \cite{bustos2020edurecomsys}, emotion detection \cite{bustos2020edurecomsys}, content-based similarity \cite{BOUSBAHI20151813}, and data mining and machine learning \cite{fotopoulou2020interactive}. We
refer the interested reader to two recent literature reviews in this area \cite{khanal2020systematic, da2023systematic}.  

While many ERSs have been proposed in the literature, there has been limited emphasis on enhancing the interactivity of these systems by incorporating various control options into the user interface (UI). Only few attempts have been made to provide interactivity and control in ERSs.
\citet{bustos2020edurecomsys} presented an interactive ERS where the user can search for educational resources based on three main criteria, namely keywords, category, and type of resource. The user can view recommendations generated through collaborative filtering or emotion detection. In the list of recommendations, the user can view more details using 'details' button. Moreover, they can provide feedback to the recommendations using five-star rating as well as mark them as favorite to view them later. Furthermore, the user can write reviews about the recommended resources in the comments section.
Another interesting attempt to interact with the ERS has been presented in \cite{santos2016toward} to offer context-aware affective educational recommendations in computer-assisted language learning in an Arduino-based platform. The recommendation module takes input through sensors and provide interactive support to learners using different communication methods like visuals, sounds, or touch. 
\citet{BOUSBAHI20151813} proposed an interactive MOOC recommender, where users can interact with the system to formulate the request as an input to the RS, e.g., add keyword for course title (text input), and select features (e.g., course fee, availability, language) using checkboxes and dropdown.
\citet{zapata2015evaluation} presented a group recommender (DELPHOS) that recommends learning objects (LOs) to a group of individuals. As an input, similar to a search engine, the user defines the desired search parameters based on a required text query or keywords, some optional metadata values and different filtering or recommendation criteria using sliders and checkboxes. Afterwards, DELPHOS shows the user a ranked list of recommended LOs which users can rate on a Likert scale of five stars, group members can add one or more tags to LOs, as well as add personal comments or additional information to them. In the area of conversational RSs in education, \citet{valtolina2024design} presented an intelligent chatbot-based RS to assist teachers in their activities by suggesting the best LOs and how to combine them according to their prerequisites and outcomes. The interaction with the RS is via text input where chatbot asks specific questions and the user provide answers in textual format. 
While there are few attempts to make ERSs more transparent by introducing
open learner models (OLMs) (e.g., \cite{barria2019explaining, abdi2020complementing}), the used OLMs, however, just show learners their system-generated interests, but do not allow learners to interact with them or modify them. 

In summary, ERSs generally provide less user control and interactivity compared to interactive RSs in other domains such as e-commerce, entertainment, and social media. A possible reason is that introducing user control in ERSs has the risk that the user
interfaces and the complexity of the interactive recommendation task might overwhelm learners, and consequently can have a negative effect on the learning experience. 
Moreover, the recent attempts to introduce interactivity with ERSs have mainly focused on providing user control mechanisms either with the input or the output of the ERS. There exists a significant research gap and potential opportunities to make ERSs more transparent by incorporating user control and interaction. To this end, in this paper, we present an interactive ERS in which we introduce different levels of user control by allowing users to interact with all the three parts of the ERS, namely input (i.e., user profile), process (i.e., recommendation algorithm), and output (i.e., recommendations). 
\subsection{Interactive Recommender Systems}
Research on RSs has traditionally focused on improving the accuracy of recommendations by developing new algorithms or integrating additional data sources into the recommendation process \cite{jannach2017user}. However, many studies have demonstrated that higher accuracy does not always enhance the user experience of the RS \cite{he2016interactive}. Consequently, recent research has shifted towards understanding how different interface elements and user characteristics impact the overall user experience with RSs. Furthermore, an effective RS should also take into account factors such as transparency to ensure societal value and trust \cite{tintarev2015explaining, harambam}. This shift in focus from purely algorithmic improvements to enhancing user experience has led to the development of what are known as interactive recommender systems (IntRS), which emphasize user control and interactivity to achieve greater transparency in RS \cite{he2016interactive, jugovac2017interacting, jannach2017user}.

IntRSs offer visual and exploratory UIs, allowing users to inspect the recommendation process and control the system to receive better recommendations \cite{he2016interactive}. These interactive, visual, and exploratory UIs progressively guide users toward their objectives, enhance their understanding of the system's functionality, and ultimately contribute to transparency \cite{he2016interactive}.
IntRSs have been developed in various domains including movies \cite{Svonava2012, schaffer2015hypothetical, loepp2014choice, Donovan, schafer2002, maxwell2015}, music \cite{2012tasteweights, saito2011, jin2018effects}, news \cite{harambam}, publications \cite{setfusion2014, bruns2015should}, tweets \cite{tintarev2015inspection}, group recommenders \cite{chen2012cofeel}, social recommenders \cite{wong2011diversity, zhao2010, smallworlds}, conference recommenders \cite{talkexplorer}, and job recommenders \cite{Donovan}. To gain a deeper understanding of IntRSs, we refer the interested reader to the excellent literature reviews on this topic in \cite{jannach2017user, he2016interactive}.

IntRSs can roughly be grouped by the level they allow users to take control on, namely the RS input (i.e., user profile), process (i.e., recommendation algorithm), and/or output (i.e., recommendations) \cite{he2016interactive, jannach2017user}. Interaction with the input of the RS allows users to create or modify their interests as they want. This level of control is provided by either allowing users to add, delete, or re-rate items in their profile using various UI elements \cite{schaffer2015hypothetical, 2012tasteweights, linkedvis2013, bruns2015should, jin2016go, schafer2002, tintarev2015inspection, jin2018effects, harambam} or allowing them to visually interact with visualizations of their interest profile to modify them \cite{Kangasrasio2015, wong2011diversity, Donovan}. Users can control the process of the RS by either choosing the recommendation algorithm \cite{Ekstrand2015LettingUC, talkexplorer, harambam}, or by manipulating the algorithmic parameters \cite{Ekstrand2015LettingUC, maxwell2015, 2012tasteweights, setfusion2014, linkedvis2013, jin2018effects}, using the UI elements provided. Lastly, users can control the output of the recommender by providing feedback to the recommendations \cite{loepp2014choice, chen2012cofeel}, or ordering and sorting the recommendations as they want \cite{2012tasteweights, harambam, maxwell2015, jin2018effects, tintarev2015inspection, talkexplorer, smallworlds, zhao2010, wong2011diversity, bruns2015should}, based on the interactive elements provided in the UI.

In summary, IntRSs offer varying levels of user interaction and control at three different levels, namely input, process, and output. As summarized in Table \ref{tab:UI design}, only a few IntRSs support interaction at all three levels \cite{2012tasteweights, saito2011, jin2018effects, harambam}. Most of the IntRSs enable user interaction at the input level, allowing users to provide or adjust their preferences. A smaller number of IntRSs facilitate interaction with the process where users can adjust algorithmic parameters. Only few recommenders allow switching between different algorithms. At the output level, most systems provide users with the ability to sort recommendations, while fewer offer options to give feedback on the recommendations. With regard to UI design, only the study in \cite{harambam} focused on the systematic design of the different control mechanisms in the UI of their proposed IntRS. To address these gaps, in this work, we introduce an interactive ERS that extends interactivity and user control which is not commonly found in the educational domain. Furthermore, we present the systematic design of the ERS in the MOOC platform CourseMapper in which we provide control across all three levels, i.e., input, process, and output of the ERS. 

The impact of user control has been investigated in the literature on interactive recommendations in various ways. Many researchers have studied the impact of user control on one or more recommendation goals, namely, perceived quality of recommendations \cite{tsai2021effects, jin2016go}, perceived accuracy of
recommendations \cite{schaffer2015hypothetical, Donovan, bruns2015should, 2012tasteweights, maxwell2015, smallworlds}, recommendation novelty \cite{loepp2014choice}, recommendation diversity \cite{wong2011diversity}, usability \cite{Kangasrasio2015, bruns2015should, zhao2010, tsai2017providing}, ease of use and playfulness \cite{chen2012cofeel}, perceived usefulness \cite{Kangasrasio2015, chen2012cofeel, tintarev2015inspection, zhao2010}, user-perceived transparency \cite{tsai2017providing, tsai2021effects}, trust in the RS \cite{schaffer2015hypothetical, bruns2015should, loepp2014choice, jin2016go, ooge23steering}, user experience with the RS \cite{knijnenburg2012inspectability, Donovan, 2012tasteweights, setfusion2014}, cognitive load and recommendation acceptance \cite{jin2018effects}, confidence with the RS \cite{schafer2002, jin2016go}, user satisfaction with the RS \cite{schaffer2015hypothetical, linkedvis2013, smallworlds, talkexplorer, tsai2017providing, jin2016go}, behavioural intentions of the user \cite{jin2016go}, and user acceptance of the RS \cite{Kangasrasio2015, linkedvis2013, jin2016go}.
While trust and 
satisfaction were the focus of a considerable number of studies, transparency remains
under-explored.
Moreover, there is a notable gap that no research has comprehensively studied the effects of user control on transparency, trust, and satisfaction  
together in the same study. Additionally, the impact of these goals on each other has yet to be explored in the interactive recommendation context. To address this research gap, in this paper, we study the impact of user control on transparency, trust, and satisfaction. Moreover, we investigate how these goals interact with each other.
\section{System Design} \label{design}
\begin{figure}[h]
  \centering
  \includegraphics[width=\linewidth]{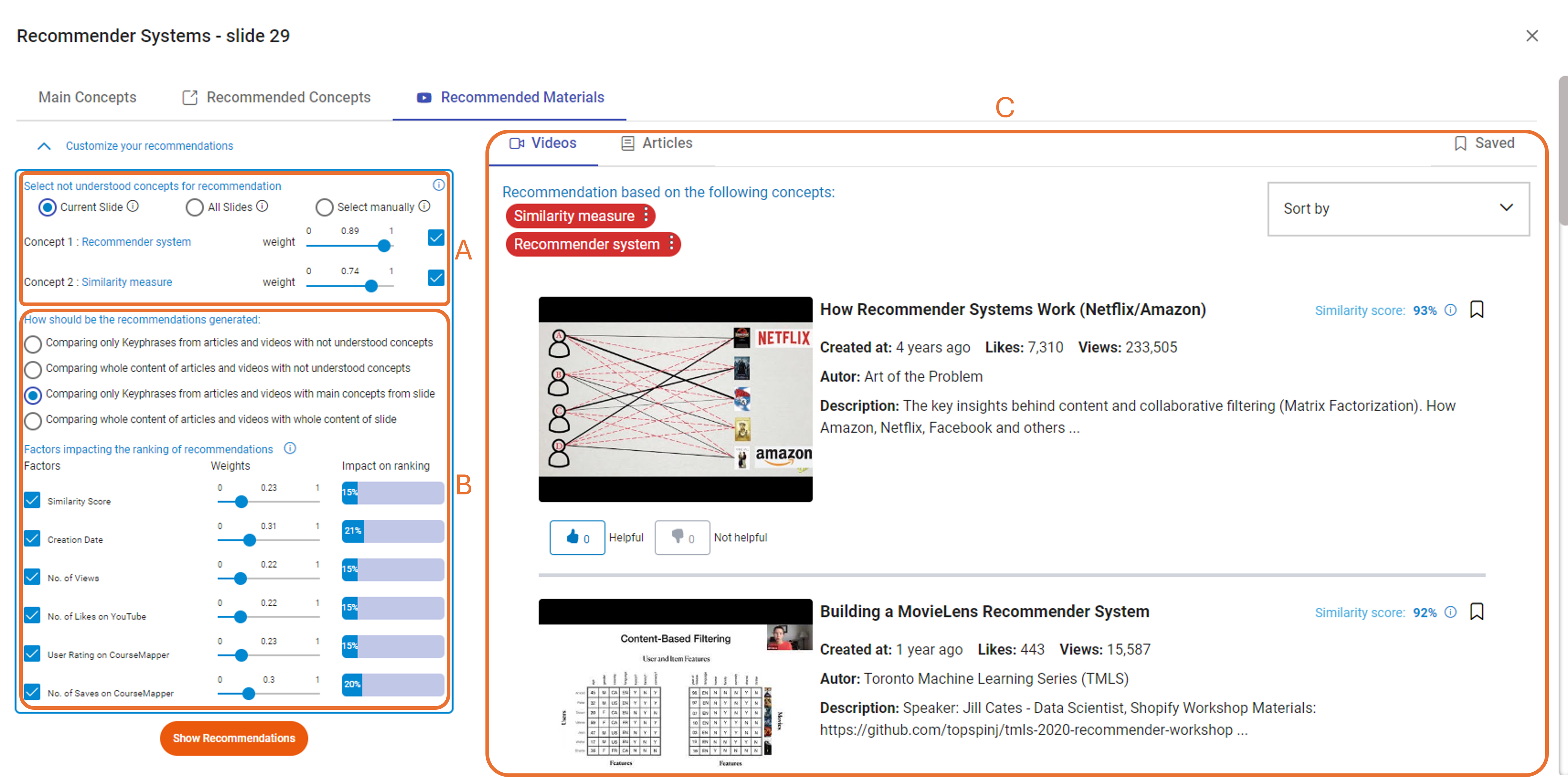}
  \caption{User Interface of the ERS in CourseMapper with three levels of user control: input (A), process (B), and output (C)}
  \label{overallinteraction}
\end{figure}

In this section, we present the design of the ERS module in the MOOC platform CourseMapper \cite{Ain2022}, which introduces user control at three different levels, namely input, process, and output (see Figure \ref{overallinteraction}). 
\subsection{User Interface Design}
In this section, we discuss the systematic approach taken to design interactive components for the UI of the ERS in CourseMapper, focusing on enhancing user control. We began by investigating the existing literature on IntRSs to identify a range of user control mechanisms and interaction options commonly employed in these systems, and analyzing their effectiveness in enhancing user control. The literature was explored focusing on answering the question: \textit{How user control has been added to the UI of the RSs to enable users to interact with different parts of the RS, i.e., input, process, and output?} Once the interaction mechanisms were identified, we chose the ones that are equally applicable to our context to ensure a better user experience. In this way, we designed interaction and control options in the UI of our ERS, focusing on the interaction with the input, process, and output of the ERS.

Beginning with the input of the IntRS, interaction with the input provides users the control to manage their preferences rather than the traditional method in which user preferences are estimated by observing their behavior over time. Many interaction options have been implemented at the input part of the recommenders to help users personalize their recommendations. Typically, to refine their requirements, users are asked to mark a set of items extracted by the system based on their past activity, using a binary scale in terms of “like/dislike” options. For instance, using ”Yes” or ”No” buttons \cite{saito2011}, or "Like" or "Dislike" buttons \cite{jin2016go}. Another option to interact with the input is provided by letting the users change the weights of their interests or re-rate items using sliders \cite{tintarev2015inspection, linkedvis2013, schaffer2015hypothetical, 2012tasteweights}, or using a pre-defined sliding scale ranging from ‘Strongly Disagree’ to ‘Strongly Agree’  \cite{wong2011diversity}. Another way is to let the users choose or modify their interests, for example, add or delete items in their profile using buttons with icons \cite{jin2018effects, schaffer2015hypothetical, linkedvis2013}, or radio buttons  \cite{tintarev2015inspection}, and re-rate items using sliders \cite{2012tasteweights, schaffer2015hypothetical, bruns2015should, harambam}. Alternatively, there are more complex methods for obtaining preferences. Examples include using filters for specific items \cite{bruns2015should}, drop down lists, checkboxes, and radio buttons to specify different dimensions of the interests \cite{jin2016go, schafer2002}, or using toggle buttons to enable/disable certain interests \cite{harambam}.
Furthermore, more advanced ways of interactions using visualizations are provided to the users using intent radar which the users can interact with using a mouse to move interest items \cite{Kangasrasio2015}, or interaction with a graph visualization of interests using a mouse to drag and drop items \cite{Donovan}.

Interaction with the process is commonly provided by allowing users to select or change the recommendation algorithm or tune the algorithm parameters. The selection of the algorithm is provided using radio buttons \cite{Ekstrand2015LettingUC}, text and icon based buttons \cite{harambam}, or selection using checkboxes \cite{talkexplorer}. To fine-tune the algorithm parameters multiple control options are provided to the users including radio buttons for feature selection \cite{saito2011}, buttons to add or remove parameters \cite{linkedvis2013}, and sliders to adjust weights of the parameters \cite{2012tasteweights, setfusion2014, bruns2015should, tsai2017providing, jin2018effects}.

Once the user’s preferences are identified, recommendation algorithm selected or modified, the system can provide tailored recommendations. Several control options and UI components have been proposed in the literature to interact with the output of the RS. Users can change the number of recommendations or filter the recommendations list using sliders \cite{Svonava2012, 2012tasteweights}, give feedback to the system using Yes/No \cite{saito2011, loepp2014choice} or Like/Dislike \cite{jin2018effects} buttons, give feedback about recommendations using buttons of different sizes referring to intensity of emotions \cite{chen2012cofeel}, or can give feedback using radio buttons \cite{maxwell2015}. Users are provided with the options to sort the recommendation list based on multiple options using buttons \cite{jin2018effects, tintarev2015inspection}, remove recommendations from the list using remove icon \cite{jin2018effects}, sort recommendations using drag and drop \cite{jin2018effects}, and reorder recommendations using toggles and sliders \cite{harambam}. Moreover, advanced interaction options are provided to the users when recommendations are presented visually instead of lists. The users can interact with the Venn diagram to examine and filter the recommended items \cite{setfusion2014}, explore the opinion space using mouse interaction to increase or decrease diameter of circular opinion space \cite{wong2011diversity}, mouse interaction with word cloud \cite{zhao2010}, drag and drop nodes in a graph \cite{smallworlds}, and arrange items in a clustermap using mouse drag interaction \cite{talkexplorer}.

\begin{figure*}[h!]
     \centering
     \begin{subfigure}{0.49\textwidth}
        
         \includegraphics[width=\linewidth]{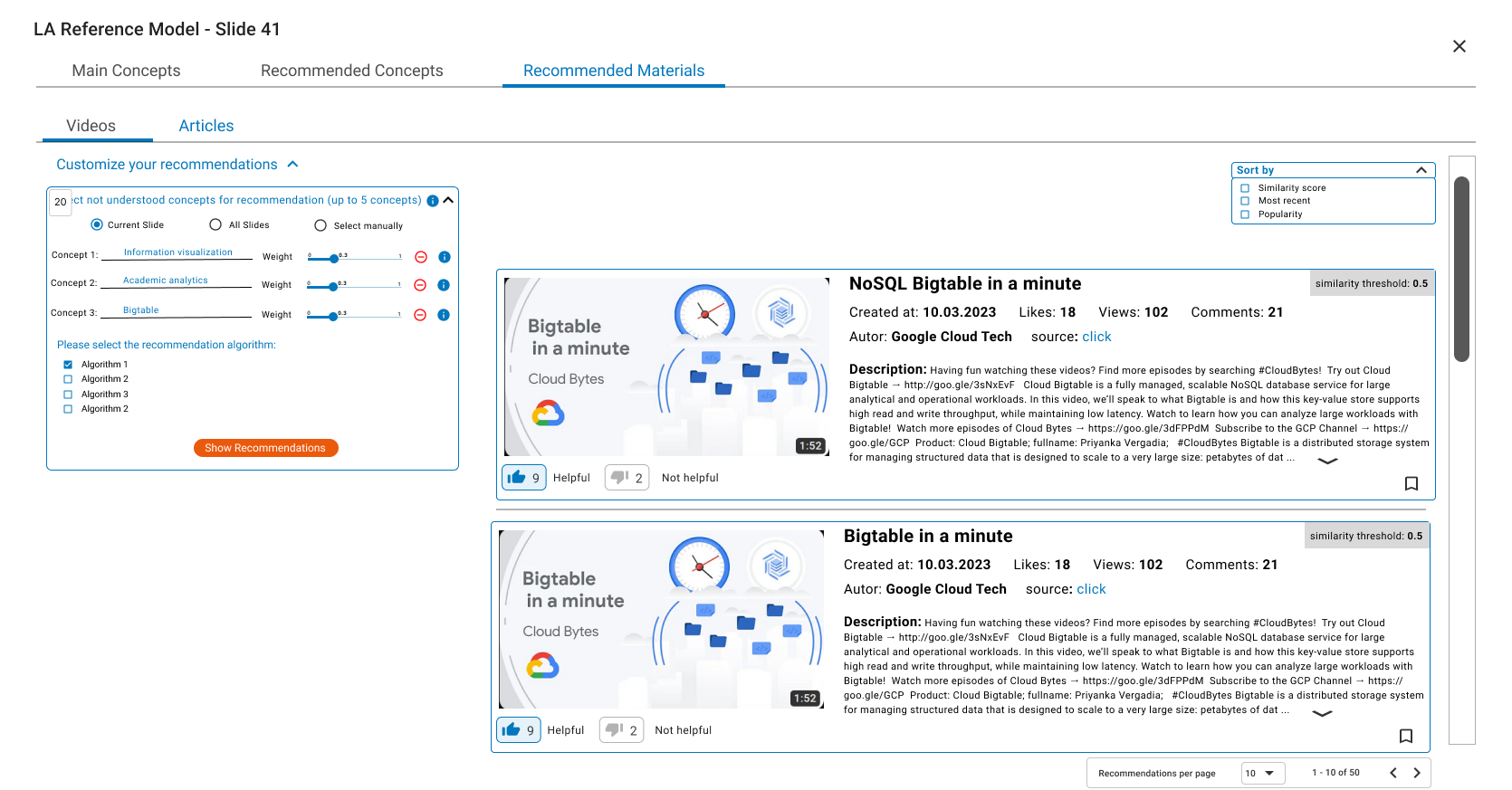}
         \caption{Initial prototype}
         \label{fig:initial}
     \end{subfigure}
     \hfill
     \begin{subfigure}{0.49\textwidth}
         \includegraphics[width=\linewidth]{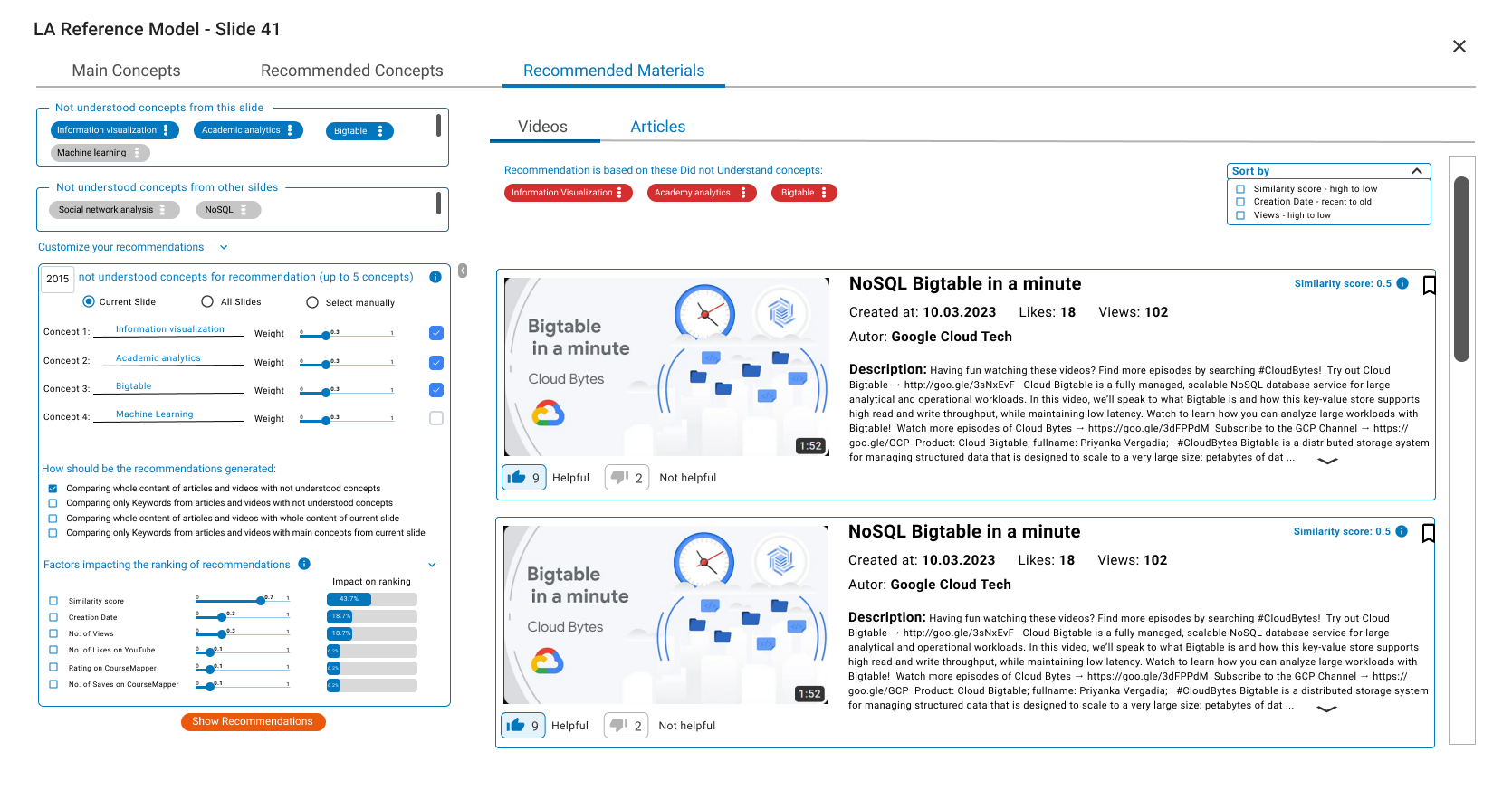}
       \caption{Improved prototype}
        \label{fig:final}
     \end{subfigure}   
        \caption{Prototypes for different levels of user control in the ERS}
        \label{fig:prototypes}
\end{figure*}

From this analysis, we identified and selected the most widely adopted interaction techniques that align with our specific context and features, ensuring that our UI design is both intuitive and effective for our users (see Table \ref{tab:UI design}). After that, we started with the design of our ERS UI. Based on the UI elements identified to interact with the input, process, and output, we created initial prototypes (see Figure \ref{fig:initial}). The prototypes were discussed with the authors' team and different UI elements were refined and improved. For example, deciding the colors, optimal labels for buttons, and labels to represent the algorithms, providing UI for ranking the recommendations, and deciding whether or not to show the impact of ranking in progress bars. The improved prototypes (see Figure \ref{fig:final}) were then translated to the final design of the system, presented in the next sections.

\begin{table}[h]
  \caption{UI design elements for interaction with different components of the RS identified from literature}
  \label{tab:UI design}
  \resizebox{\textwidth}{!}{%
    \begin{tabular}{lp{4cm}p{4cm}p{5cm}} 
      \toprule
      Paper & User control with input & User control with process & User control with output\\
      \midrule
      \citet{jin2016go} & Like/Dislike buttons, \newline Selection using radio buttons, and dropdowns & - & - \\
      \citet{schafer2002} & Selection using radio buttons, checkboxes, \newline and dropdowns & - & - \\
      \citet{schaffer2015hypothetical} & Add or delete buttons, \newline Sliders to re-rate items & - & - \\
      \citet{Kangasrasio2015} & Mouse interaction with radar & - & - \\
      \citet{Donovan} & Mouse interaction with graph & - & - \\
      \citet{linkedvis2013} & Sliders to adjust weights & Buttons to add or remove parameters & - \\
      \citet{bruns2015should} & Filter items using buttons & Sliders to adjust weights & - \\
      \citet{2012tasteweights} & Sliders to adjust weights & Sliders to adjust weights & Filter using sliders \\
      \citet{saito2011} & Yes/No buttons & Feature selection using radio buttons & Feedback using Yes/No buttons \\
      \citet{loepp2014choice} & - & - & Feedback using Yes/No buttons \\
      \citet{Svonava2012} & - & - & Filter using sliders \\
      \citet{jin2018effects} & Select input using buttons & Sliders to adjust weights & Feedback using Like/Dislike buttons, drag to sort, remove item \\
      \citet{maxwell2015} & - & - & Feedback using radio buttons \\
      \citet{chen2012cofeel} & - & - & Feedback using size of buttons \\
      \citet{harambam} & Toggle to enable/disable interests & Select using text buttons & Reorder using toggle \\
      \citet{tintarev2015inspection} & Sliders to adjust weights & - & Sort using buttons \\
      \citet{setfusion2014} & - & Sliders to adjust weights & Filter in Venn diagram \\
      \citet{wong2011diversity} & Fixed sliders to adjust weights & - & Mouse interaction with circular opinion space \\
      \citet{zhao2010} & - & - & Mouse interaction with word cloud \\
      \citet{smallworlds} & - & - & Drag and drop nodes in a graph \\
      \citet{talkexplorer} & - & Select using checkboxes & Drag and drop in cluster map \\
      \citet{tsai2017providing} & - & Sliders to adjust weights & - \\
      \citet{Ekstrand2015LettingUC} & - & Select using radio buttons & - \\
      \textbf{ERS in CourseMapper} & \textbf{Add, delete, select using buttons, include/exclude using checkboxes, adjust the weights using sliders} & \textbf{Select using radio buttons, adjust weights for ranking using sliders, view impact of adjustments in progress bars} & \textbf{Save using buttons, Sort using dropdown options, Feedback with connected input selection using Helpful/Not Helpful button}\\
      \bottomrule
    \end{tabular}
  }%
\end{table}

\subsection{Interaction with the Recommendation Input}
\begin{figure*}[t!]
     \centering
     \begin{subfigure}{0.3\textwidth}
         \centering
         \includegraphics[width=\linewidth]{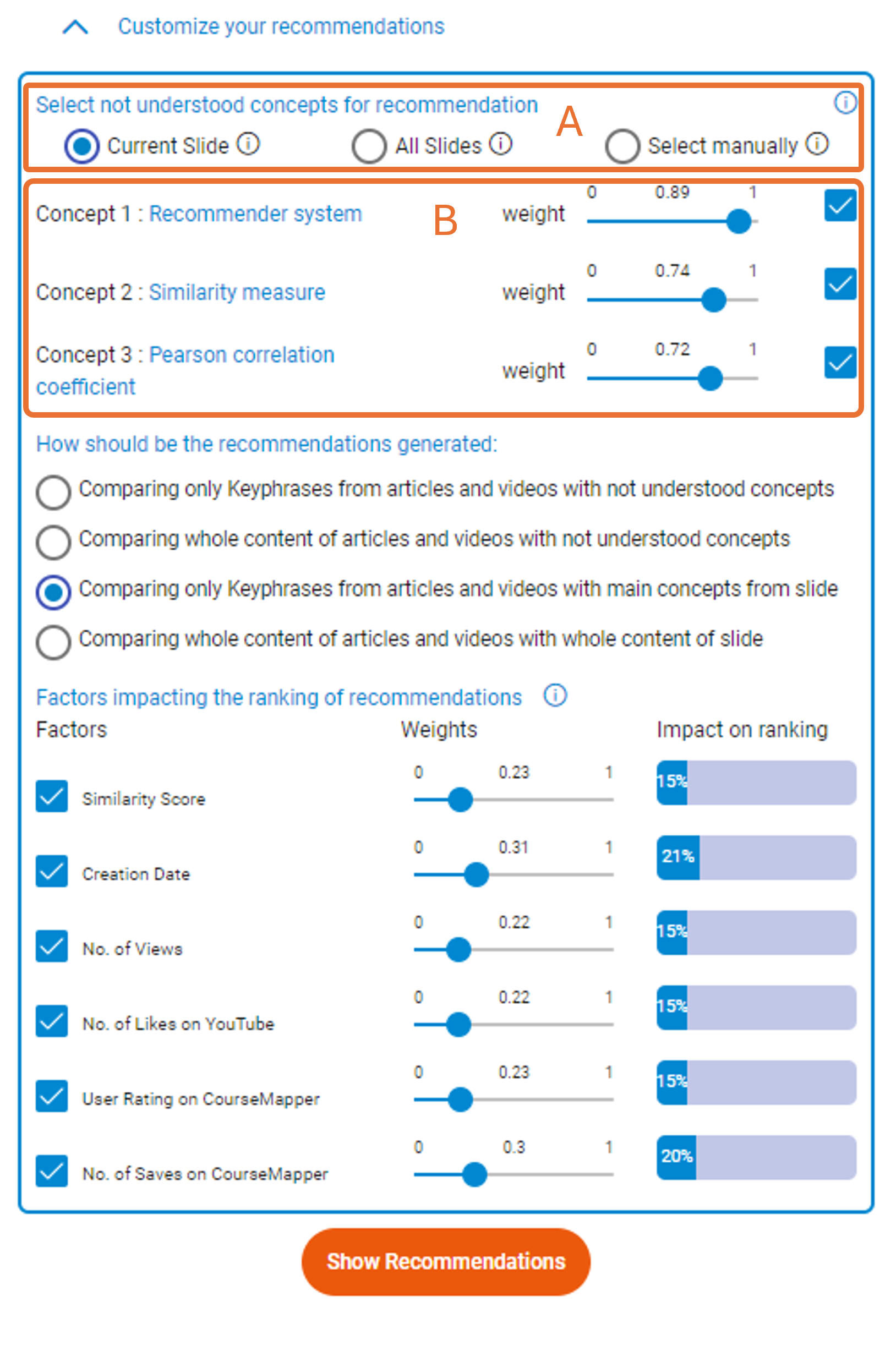}
         \caption{Select DNU concepts from the current slide as input for recommendation}
         \label{fig:current}
     \end{subfigure}
     \hfill
     \begin{subfigure}{0.3\textwidth}
         \centering
         \includegraphics[width=\linewidth]{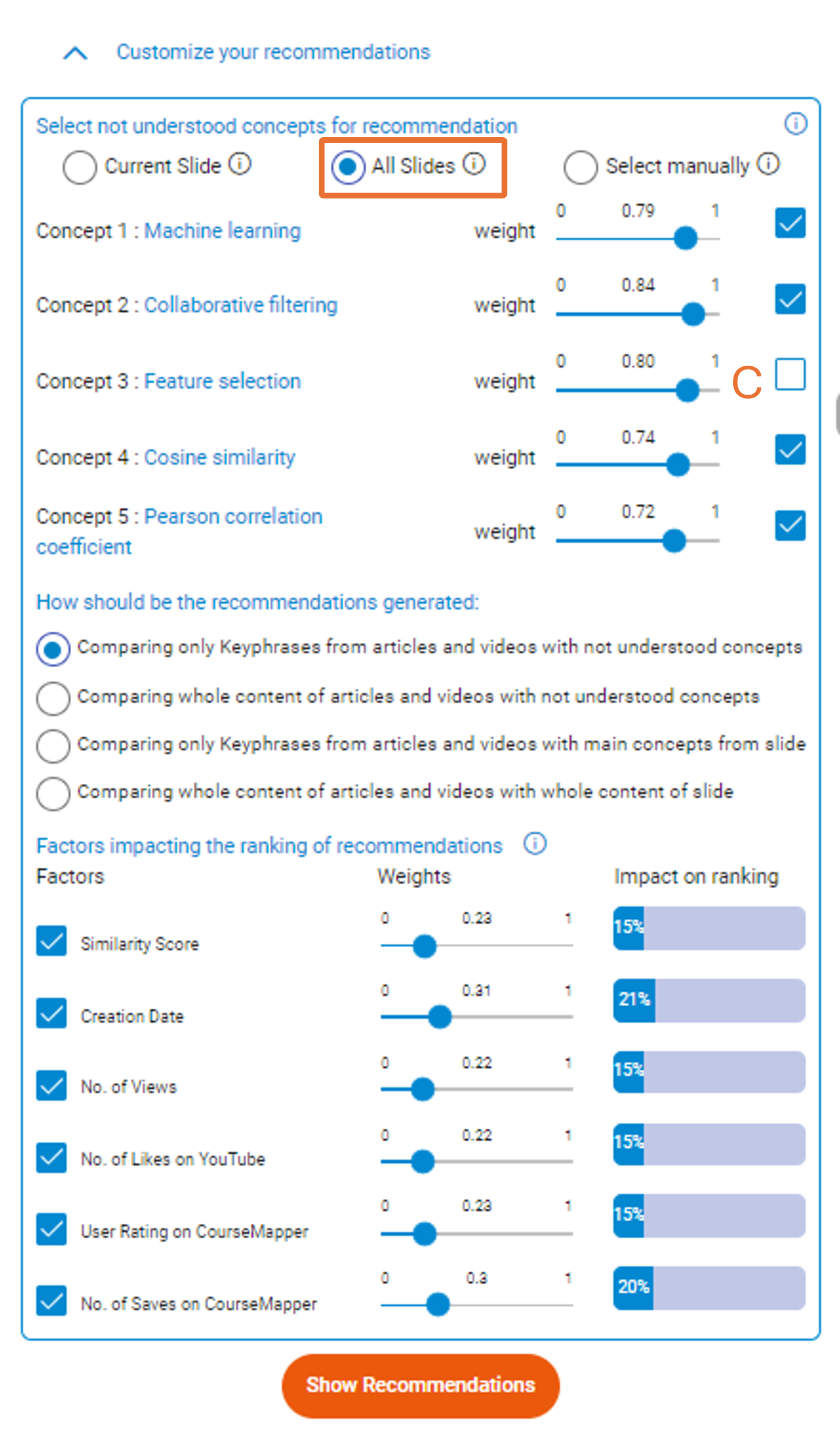}
       \caption{Select DNU concepts from all slides as input for recommendation}
        \label{fig:allslides}
     \end{subfigure}
          \hfill
     \begin{subfigure}{0.3\textwidth}
         \centering
         \includegraphics[width=\linewidth]{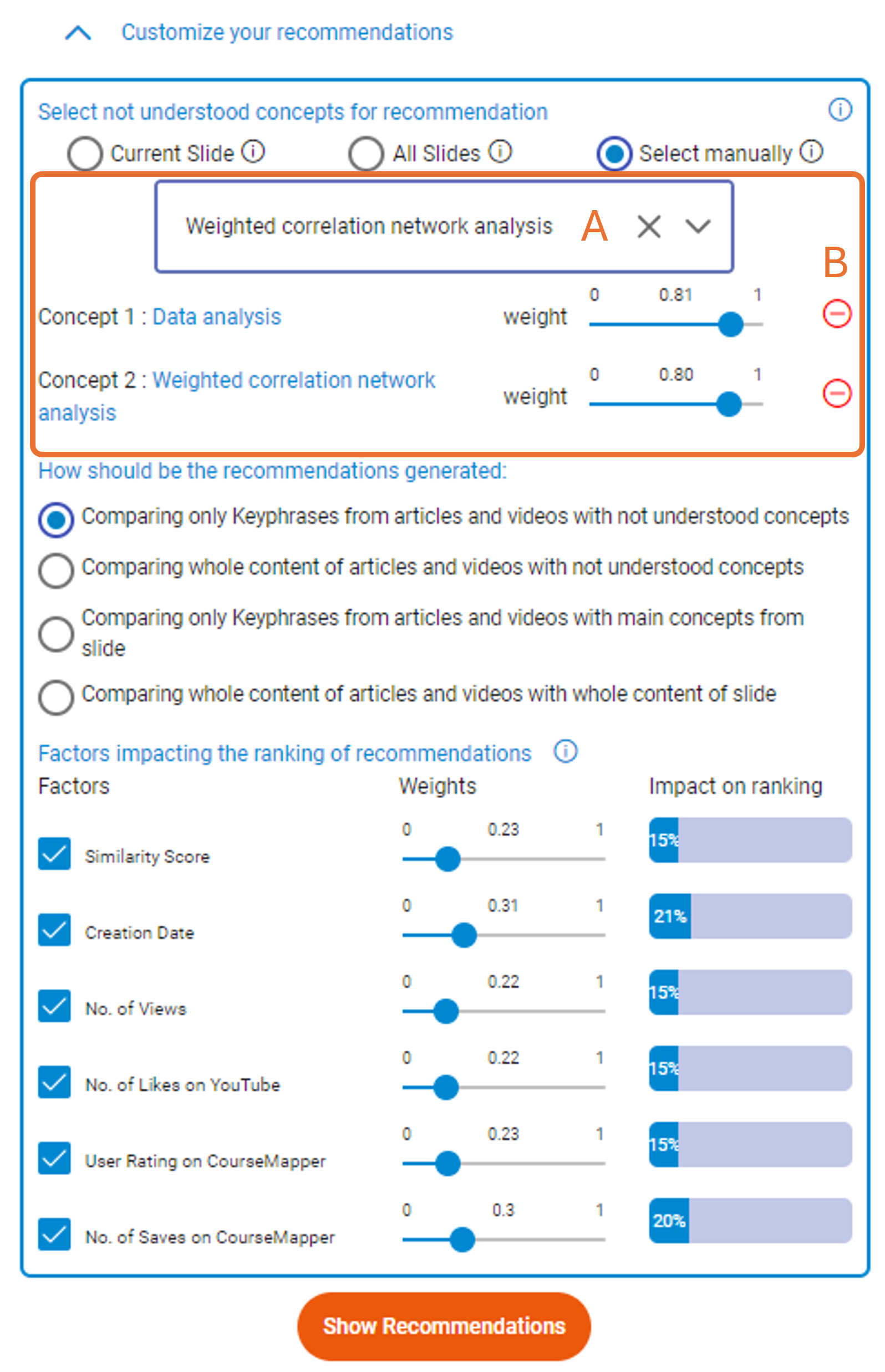}
       \caption{Manually select concepts from the learning material as input for recommendation}
        \label{fig:manually}
     \end{subfigure}
        \caption{Interaction with the input of the ERS}
        \label{fig:input}
\end{figure*}

In the context of recommending learning resources, it is crucial to recommend accurate resources tailored to learners’ needs for
better learning outcomes \cite{ain2024learner}. Therefore, providing interaction around the input of the ERS will facilitate learners to directly communicate their interests to the system. A simple way to give users more control is to allow them to explicitly specify their interests and preferences rather than relying on the system to determine these preferences from their past interactions \cite{jannach2017user}. Similarly, in our ERS, to facilitate interaction with the input, we provide control to the users to choose their interests and preferences explicitly to construct their learner model. Since CourseMapper \cite{Ain2022} is a MOOC platform where learners can enroll into multiple courses which contain learning materials, the learners have the option to decide their knowledge deficiencies by explicitly marking the concepts as 'Did Not Understand' (DNU concepts) from various slides of the PDF learning materials. To proceed with recommendations the learners have various control options to decide the input for recommender. Figure \ref{fig:input} shows the UI of the ERS with multiple options to interact with the input. First, the users can decide which DNU concepts they want the recommendations for, using the radio buttons to select 'Current Slide', 'All Slides', or 'Select manually' options (Figure \ref{fig:current} (A)). Clicking on 'Current Slide' and 'All Slides' retrieves the DNU concepts related to the current slide or all slides of the learning material, respectively (Figure \ref{fig:current} (B), \ref{fig:allslides}). Whereas, clicking on the 'Select manually' option let the user choose any concepts from the whole learning material using the dropdown list with a search option (Figure \ref{fig:manually} (A)). Once the DNUs are selected as input for the recommender, in all the three options, the user is provided the opportunity to manipulate the weights of the concepts based on their preferences if they find some concepts more important than others (Figure \ref{fig:current} (B)). The recommendations will be impacted by the changes in weights correspondingly. Furthermore, the user can include/exclude certain DNUs using checkboxes (Figure \ref{fig:allslides} (C)), as well as remove DNUs from the input list using '-' icon (Figure \ref{fig:manually} (B)).

\subsection{Interaction with the Recommendation Process}
Interacting with the recommendation process allows users to choose or influence the recommendation strategy or algorithm parameters \cite{jannach2019explanations}. 
To introduce user control around the process of our ERS, we provide two options to the user. The first is to let them choose between the four recommendation algorithms using radio buttons (Figure \ref{fig:process}, (A)). For better understandability of the recommendation process for users, we provide a text-based abstract description of the underlying algorithms. The options to select the algorithm are presented as an answer to the question "How should the recommendations be generated?", with four possible answers as: 1) Comparing only keyphrases from articles and videos with DNU concepts, 2) Comparing the whole content of articles and videos with DNU concepts, 3) Comparing only keyphrases from articles and videos with main concepts of the current slide, and 4) Comparing the whole content of articles and videos with whole content of the current slide. For more details about how these recommendation algorithms work, please refer to our earlier publication \cite{ain2024learner}. Once the user selects the algorithm, the second control option is to decide how they want the recommendations to be ranked. The user can view multiple factors impacting the ranking of the recommendations with their default weights. The users can adjust the weights of the factors using sliders (Figure \ref{fig:process}, (B)), where the impact of the change in weight on the ranking is displayed in real-time in the progress bar adjacent to each factor (Figure \ref{fig:process}, (C)). Furthermore, the user can select/de-select factors from the list using checkboxes if they do not want to include a factor in ranking (Figure \ref{fig:process}, (D)). 

\subsection{Interaction with the Recommendation Output}
\begin{figure*}[h]
    \centering
    \begin{subfigure}{0.32\textwidth}
        \includegraphics[width=\linewidth]{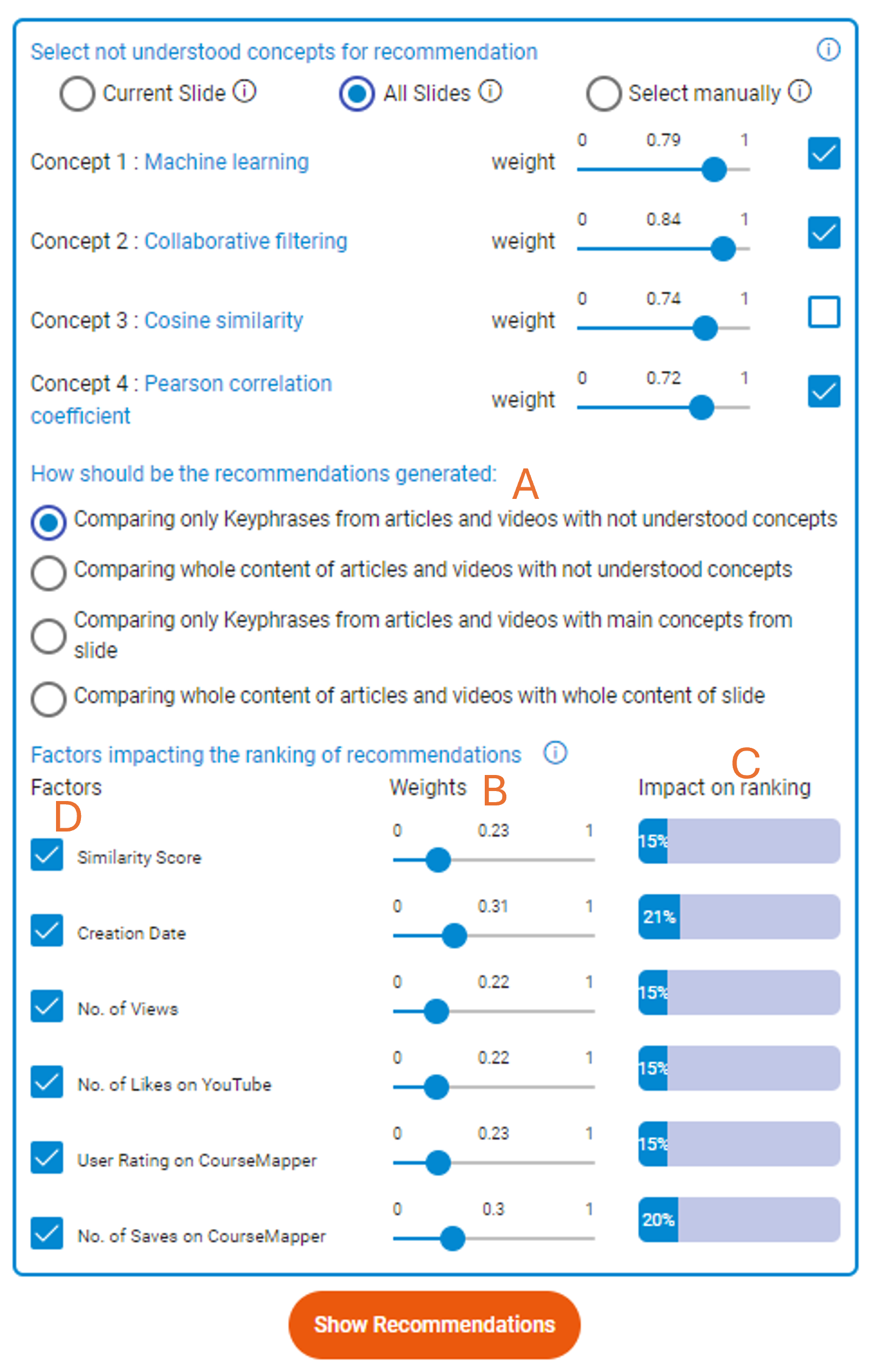}
        \caption{Interaction with the process of the ERS}
        \label{fig:process}
    \end{subfigure}
    \hfill
    \begin{subfigure}{0.67\textwidth}
        \raisebox{15mm}{ 
            \includegraphics[width=\linewidth]{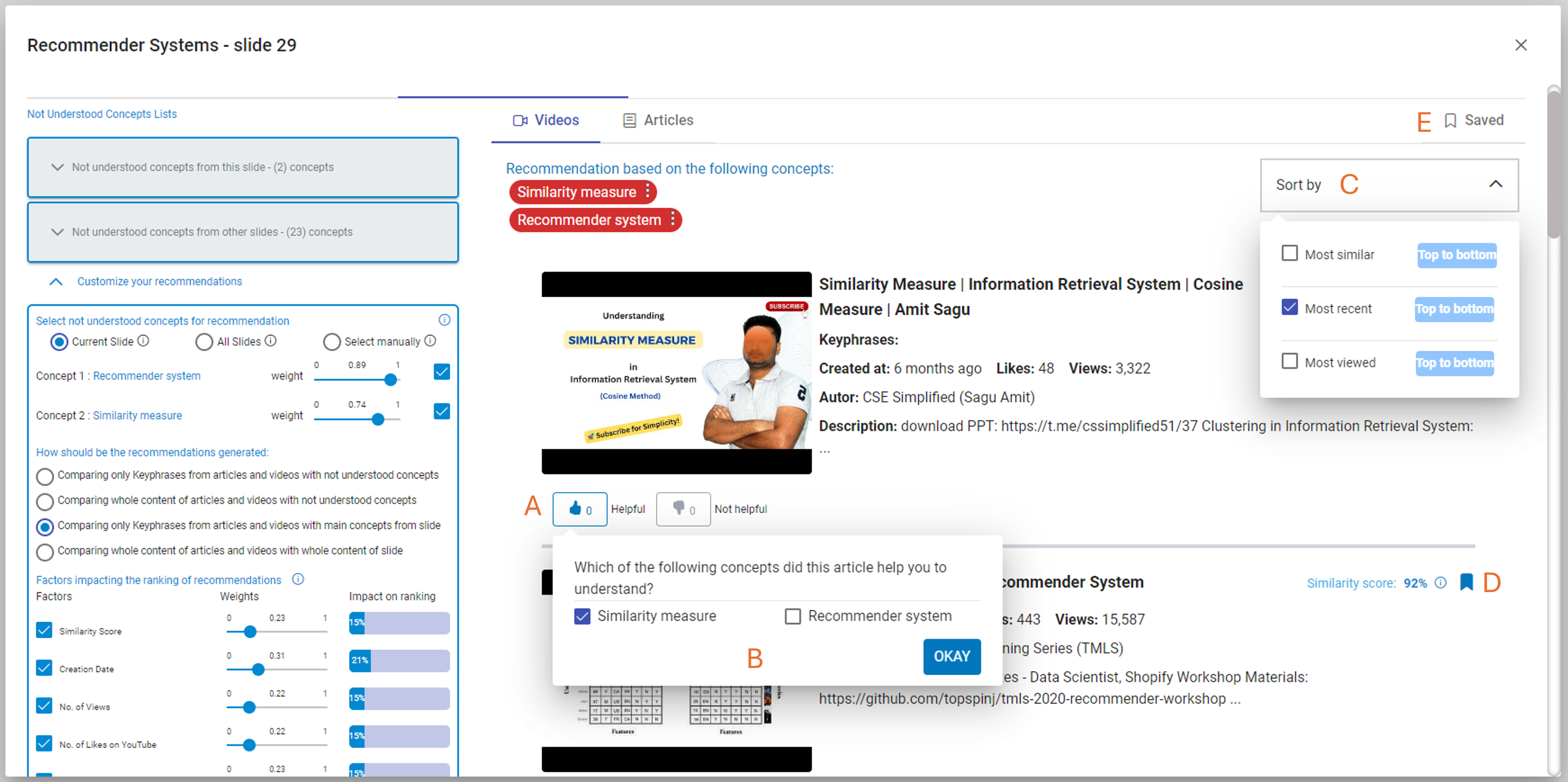}
        }
        \caption{Interaction with the output of the ERS}
        \label{fig:output}
    \end{subfigure}   
    \caption{Interaction with the process and output of the ERS}
    \label{fig:prout}
\end{figure*}

Providing feedback to the system about the provided recommendations is an important factor influencing the user experience of the RS \cite{schaffer2015hypothetical}. We provide multiple control options to the user to interact with the output of the ERS (see Figure \ref{fig:output}). The first one is to let users provide feedback to the system about the provided recommendations. To facilitate this, we provide a 'Helpful/Not Helpful' button with each recommended video or article (Figure \ref{fig:output}, (A)). Once the user clicks on 'Helpful' a dropdown menu appears which prompts the user to select the DNU concepts that the recommended video or article helped them understand (Figure \ref{fig:output}, (B)). This extended level of control enables the user to provide detailed feedback to the ERS, which can be used to improve future recommendations. Furthermore, the user can sort the recommendations using the sort option which opens a dropdown with multiple options (Figure \ref{fig:output}, (C)). The user can sort the recommendations based on their similarity score, creation date, or number of views. The options are presented as 'Most similar', 'Most recent' and 'Most viewed'. The third interaction option is to save the recommendations using a save icon (Figure \ref{fig:output}, (D)). These recommendations are saved to access them later in the 'Saved' tab (Figure \ref{fig:output}, (E)).

\section{User Evaluation} \label{evaluation}
To evaluate our system, we conducted a detailed user study with end users, employing key measures from the ResQue evaluation framework \cite{Pu2011resque} to evaluate users’ perceived
benefits in terms of perceived system qualities (recommendation accuracy, recommendation novelty, recommendation diversity, interface adequacy, information sufficiency, interaction adequacy), beliefs (perceived ease of use, control, transparency, perceived usefulness), attitudes (overall satisfaction, confidence, trust), and behavioral intentions (use intentions, watch intention).

\subsection{Participants and Procedure}
We recruited participants from our current and past courses via emails and phone contacts. In total, 30 students (5F, 25M) participated in the study, including 10 Bachelor, 14 Masters
and 6 Ph.D., from various age groups (21 from 20-30, 8 from 30-40 and 1 above 40 years) and countries (6 Cameroon, 2 Chineese, 6 Egyptian, 2 German, 1 Nigerian, 4 Pakistani, 7 Syrian, and 2 Tunisian). Participants were from various
backgrounds including computer science and applied computer science, physics, and engineering. Most of the participants were familiar with the use of RSs (n=24, 63\%) and interacting with RSs (n=19, 63\%). However, only a few of them (n=12, 40\%) were familiar with technical implementations of RSs.

The study was conducted online through individual sessions with each participant using video conferencing on Zoom. At the start of each session, participants were informed about the voluntary nature of their participation, the confidentiality and anonymity of data collection, and were asked for their consent to record the meeting. Following this, they completed a demographic survey and were then introduced to the ERS via a demo video. Afterwards, participants were asked to perform specific tasks using the ERS by taking control of the host's shared screen. Once done, they filled an online questionnaire using Google forms based on measures from ResQue \cite{Pu2011resque}. 
The study took 90 minutes on average.

\subsection{User Tasks}
The main tasks for the participants were to interact with multiple control options provided with the input, process, and output of the ERS. The tasks were divided into sub-tasks to distribute the cognitive load and facilitate a smooth experience with the ERS.
\begin{enumerate}[label=\arabic*.]
  \item Task 1: Interaction with the ERS input
    \begin{enumerate}[label=\alph*.]
      \item Imagine that you are reading a learning material in CourseMapper and while reading you do not understand something from the slides. Please interact with some slides and collect the concepts you do not understand. \textit{(Goal: to add items to their interests)}.
      \item Now, you want to view the recommended YouTube videos based on the specific concepts that you do not understand from a particular slide. How will you proceed with it using this system? \textit{(Goal: to test 'current slide' option).}
      \item Once you have viewed the provided recommendations, now you want to generate new recommendations based on some other concepts that you did not understand. Moreover, you feel that some concepts are more important than the other ones and should be included with a higher importance while generating the recommendations, or that some concepts should not be included in the recommendations. How will you proceed with it? \textit{(Goal: to test 'all slides/select manually', weight adjustment using sliders, include/exclude using checkboxes, and remove options).}
       \end{enumerate}
    \item Task 2: Interaction with the ERS process
    \begin{enumerate}[label=\alph*.]
    \item In this recommender system, you have the option to decide the way you want the recommendations to be generated. There are multiple recommendation algorithms that you can select based on your requirements. How would you proceed with changing the recommendation algorithm as per your choice?  \textit{(Goal: to test 'select algorithm' option).}
    \item The recommendations are ranked after generation in the system’s default way. Imagine that you want the recommendations to be ranked differently as you prefer. How would you proceed? \textit{(Goal: to test 'modify ranking' options).}
    \end{enumerate}
  \item Task 3: Interaction with the ERS output
  \begin{enumerate}[label=\alph*.]
  \item Now, you have the videos recommended to you and you want to view the latest videos first. Imagine that you don’t have time to watch all of the recommended videos now, and then ensure that you can easily access a few of them later when you have more time to watch them.  How will you proceed? How will you access them later? \textit{(Goal: to test 'sort' and 'save' options).}
   \item Imagine that you find some of the recommended videos more useful than others in understanding the concepts you previously did not understand. How will you give this feedback to the system? \textit{(Goal: to test 'Helpful/Not helpful' option).}
   \end{enumerate}
\end{enumerate}

\section{Results} \label{results}
We asked all study participants to fill in a questionnaire after their
interaction with the ERS. This questionnaire assessed their experience of controlling different parts of the ERS on
a five-point Likert scale, using items from the ResQue evaluation framework \cite{Pu2011resque}. Table \ref{resque} lists the items and the results based on data obtained from the questionnaires. 

\begin{table}[htbp]
\caption{Post study questionnaire and results based on ResQue \cite{Pu2011resque}}
\label{resque}
\centering
\resizebox{\textwidth}{!}{%
\begin{tabular}{lll}
\toprule
Measure                                & Questions                                                                                                  & Score (Mean (SD))             \\
\midrule
Recommendation Accuracy                & The items recommended to me matched my interests.                                                          & 4.03 (0.75)                  \\
\midrule
Recommendation Novelty                 & The recommender system helped me discover new videos.                                                      & 4.13 (0.56)                  \\
\midrule
Recommendation Diversity               & The items recommended to me are diverse.                                                                   & 3.86 (0.56)                  \\
\midrule
\multirow{4}{*}{Interface Adequacy}    & The labels of the recommender interface are clear.                                                         & \multirow{4}{*}{3.86 (0.7)} \\
                                       & The labels of the recommender interface are adequate.                                                      &                    \\
                                       & The layout of the recommender interface is attractive.                                                     &                    \\
                                       & The layout of the recommender interface is adequate.                                                       &                    \\
                                       \midrule
Information Sufficiency                & The information provided for the recommended videos is sufficient for me to make a decision to watch them. & 4.33 (0.59)                  \\
\midrule
\multirow{2}{*}{Interaction Adequacy}  & I found it easy to tell the system what I like/ dislike.                                                   & \multirow{2}{*}{4.45 (0.75)} \\
                                       & I found it easy to inform the system if I like/dislike the recommended item.                               &                    \\
                                       \midrule
\multirow{2}{*}{Perceived Ease of Use} & I became familiar with the recommender system very quickly.                                                & \multirow{2}{*}{4.13 (0.79)} \\
                                       & I easily found the recommended items.                                                                      &                    \\
                                       \midrule
\multirow{2}{*}{Control}               & I feel in control of modifying my interest profile.                                                        & \multirow{2}{*}{4.16 (0.8)} \\
                                       & I found it easy to modify my interest profile in the recommender.                                          &                    \\
                                       \midrule
Transparency                           & I understood why the videos were recommended to me.                                                        & 4.2 (0.7)                  \\
\midrule
\multirow{3}{*}{Perceived Usefulness}  & The recommender helped me find the ideal item.                                                             & \multirow{3}{*}{3.94 (0.74)} \\
                                       & Using the recommender to find what I like is easy.                                                         &                    \\
                                       & The recommender gave me good suggestions.                                                                  &                    \\
                                       \midrule
Overall Satisfaction                   & Overall, I am satisfied with the recommender.                                                              & 4.26 (0.62)                  \\
\midrule
\multirow{2}{*}{Confidence}            & I am confident I will like the items recommended to me.                                                    & \multirow{2}{*}{4.05 (0.92)} \\
                                       & The recommender made me more confident about my selection/decision.                                       &                    \\
                                       \midrule
Trust                                  & The recommender can be trusted.                                                                            & 3.9 (0.9)                  \\
\midrule
\multirow{3}{*}{Use Intentions}        & I will use this recommender again.                                                                         & \multirow{3}{*}{4.04 (0.88)} \\
                                       & I will use this recommender frequently.                                                                    &                    \\
                                       & I will tell my friends about this recommender.                                                             &                    \\
                                       \midrule
Watch Intention                        & I would watch the videos recommended, given the opportunity.                                               & 4.3 (0.58)    \\
\bottomrule
\end{tabular}%
}
\end{table}

\begin{figure}[h]
  \centering
  \includegraphics[width=0.9\linewidth]{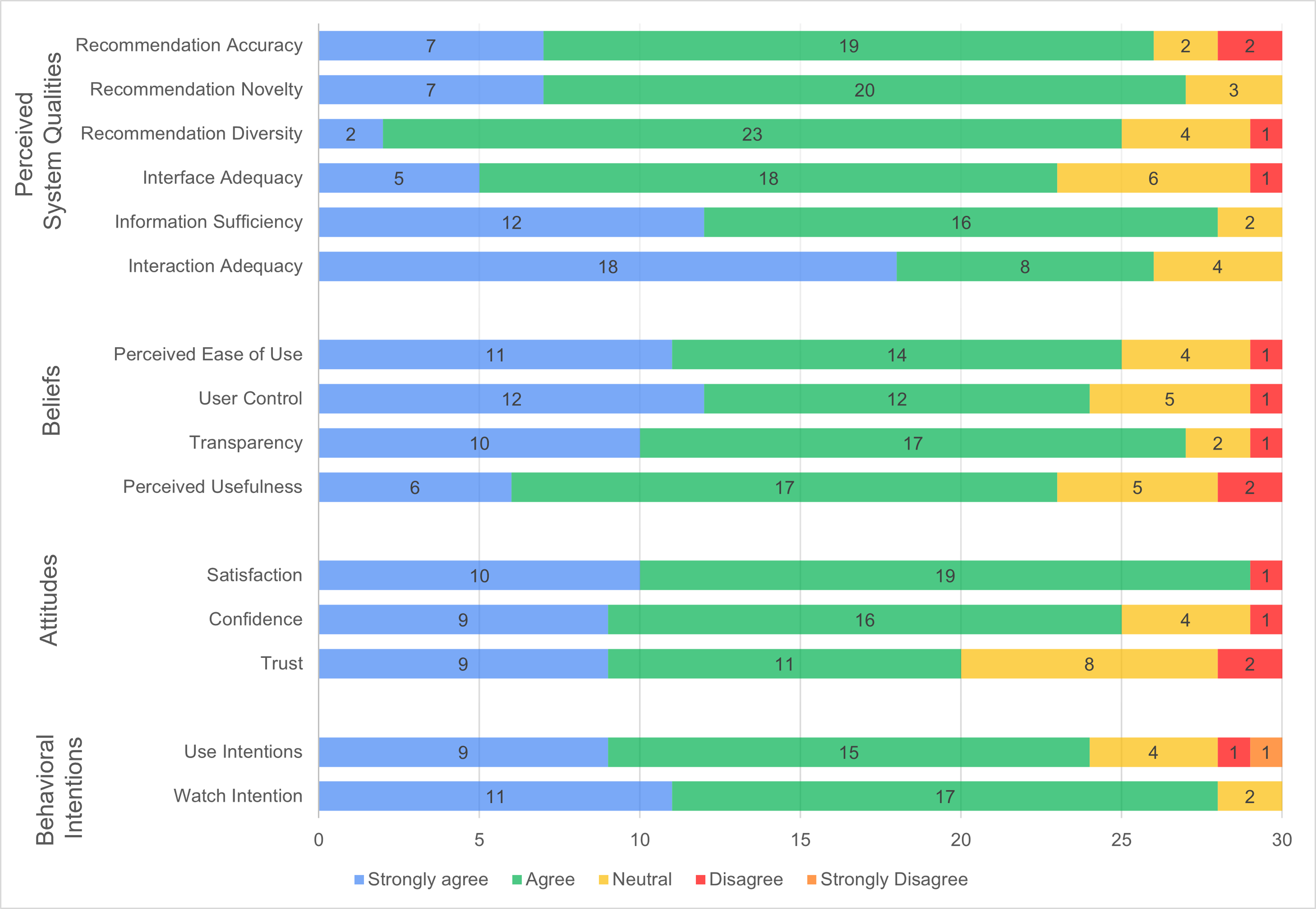}
  \caption{Results of user evaluation based on ResQue \cite{Pu2011resque}}
  \label{result}
\end{figure}

\subsection{Perceived Benefits for Users}
Concerning RQ1 ("How does complementing an ERS with user control impact users' perceptions of the ERS?"), the analysis of the questionnaire data shows that users' feedback on the ERS is positive across all perceived aspects. 
Regarding recommendation-related perceived system qualities, the scores for recommendation accuracy (Mean=4.03, SD=0.75) and recommendation novelty (Mean=4.13, SD=0.56) are relatively high, suggesting that user control over the ERS can lead to relevant and novel recommendations that align with users' interests and preferences. Similarly, for UI-related perceived system qualities, the scores for information sufficiency (Mean=4.33, SD=0.59) and interaction adequacy (Mean=4.45, SD=0.75) are also high, with interaction adequacy scoring the highest among all measures. This indicates that users appreciated the interaction with the ERS input, process, and output and found it easy to communicate their preferences and feedback to the ERS. Compared to information sufficiency and interaction adequacy, interface adequacy received a lower score (Mean=3.86, SD=0.7), indicating that the interface design needs improvements in terms of labels and layout.   

Regarding user beliefs which refers to a higher level of
user perception of a system, which is influenced by perceived
qualities \cite{Pu2011resque}, the score for user control (Mean=4.16, SD=0.8) indicates that users felt in control of their interactions with the ERS. Additionally, a high score for transparency (Mean=4.2, SD=0.7) suggests that the ability to control different parts of the ERS helped users understand how the ERS works. These high scores suggest that user control is positively associated with transparency.  
We explore in detail the impact of user control on transparency and other recommendation goals in Section \ref{impact}. Furthermore, the score for perceived ease of use (Mean=4.13, SD=0.79) suggests that the ERS facilitates users to find their preferred items quickly and that users found the ERS easy to navigate and quickly became familiar with it. In contrast, perceived usefulness received a comparatively lower score (Mean=3.94, SD=0.74), possibly because the diverse recommendations were not always perceived as optimal suggestions for some participants. 

With regard to user attitudes referring to a user’s overall feeling towards an RS \cite{Pu2011resque}, a highly positive score of confidence (Mean=4.05, SD=0.92) reflects the recommender's ability to convince users of the recommended items. 
Furthermore, users showed a high level of overall satisfaction with the ERS (Mean=4.26, SD=0.62), indicating an increased ease of use and enjoyment with the ERS.  
According to \citet{tintarev2007}, satisfaction can also be measured indirectly, measuring user loyalty. Thus, users’ behavioral intentions can be seen as an indirect measure of loyalty and satisfaction with the system. To this end, the highly positive scores for use intentions (Mean=4.04, SD=0.88) and watch intention (Mean=4.3, SD=0.58) suggest that the system effectively influenced users' decisions to continue using the ERS and reflects their satisfaction with the ERS. In terms of trust, the positive score (Mean=3.9, SD=0.9) shows an increased users' confidence in the ERS. 
However, trust scored comparatively less than transparency and satisfaction. This suggests that, while the recommendation goals of transparency, trust, and satisfaction appear to move together, trust seems to be less correlated with the other two goals. We investigate in detail the relationships and correlations between these goals in Section \ref{goals}.
\begin{figure} [h]
  \centering
  \includegraphics[width=0.6\linewidth]{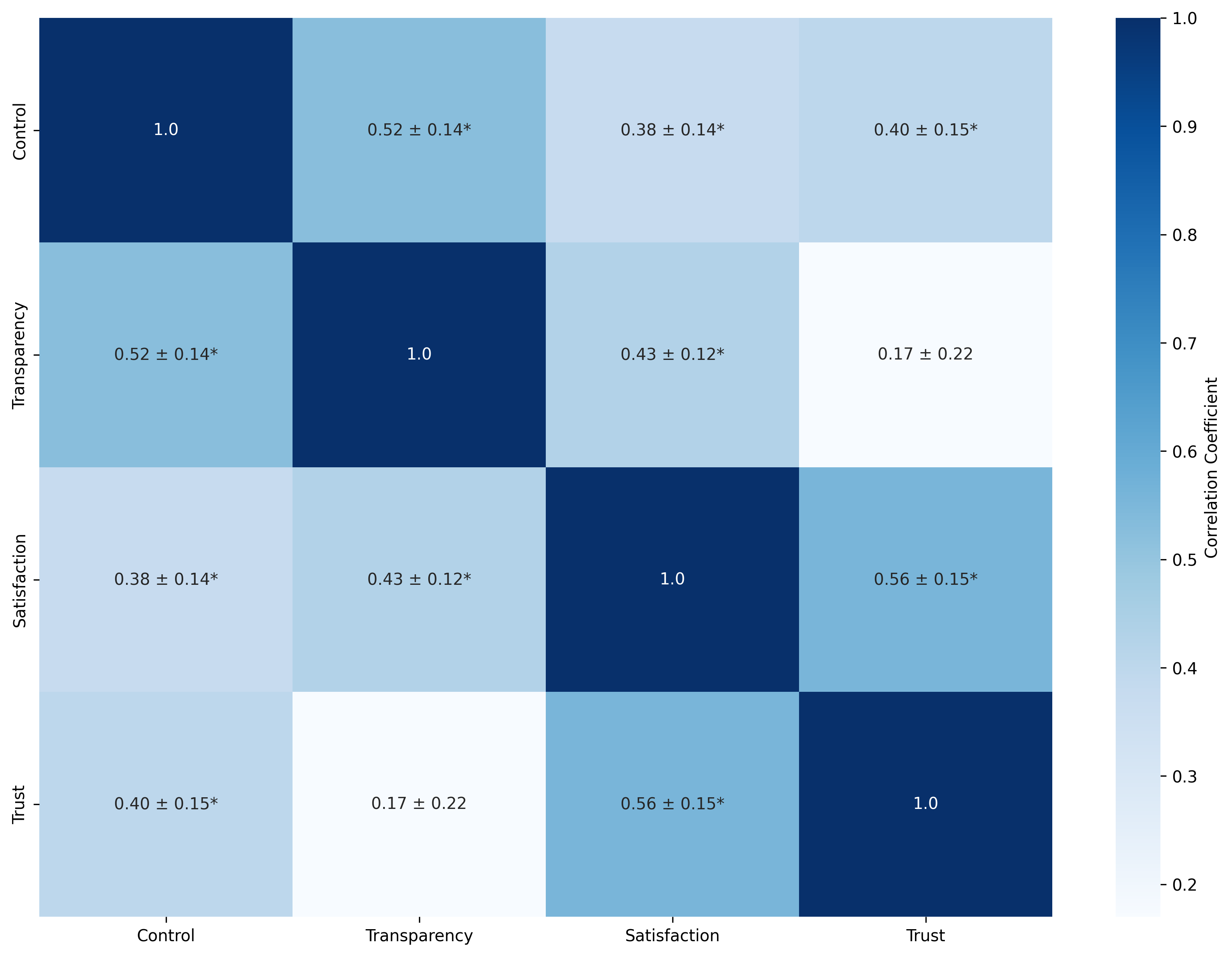}
  \caption{Pearson Correlation between different goals, with 95\% confidence interval using bootstrap sampling, where statistically significant correlations (adjusted p-value < 0.05) are marked with an asterisk (*)}.
  \label{correlation}
\end{figure}
\subsection{Impact of User Control on Different Goals} \label{impact}
To identify the impact of user control on different recommendation goals, namely transparency, trust, and satisfaction (RQ2), and to explore how these goals interact with each other (RQ3), we computed Pearson Correlation between them. Figure \ref{correlation} shows the Pearson Correlation between each pair of goals, by analyzing the responses of participants from the user study. For each correlation, the figure also shows a 95\% confidence interval for the correlation measured using bootstrap sampling, as well as if the correlations are statistically significant or not (adjusted p-value < 0.05). The results from the correlation analysis reveal that user control has at least moderate correlation with all goals. More specifically, user control strongly correlates with transparency with a statistical significance. Therefore, our study provides evidence that user control with the ERS leads to an increased transparency. This confirms findings from previous studies from the recommendation domain which pointed out that user control is very closely tied to  transparency \cite{tintarev2015explaining} and that control significantly affects transparency \cite{Pu2011resque}. This is also in line with studies from interactive recommendation literature which show that user control can also contribute to increased transparency of RSs \cite{tsai2021effects, tsai2017providing}. 
Furthermore, in our study, most users (n=27, 90\%) regardless of their background knowledge concurred that the ERS was perceived as transparent. One reason we deem responsible for the high level of transparency among the majority of participants is that they were able to control the ERS in a way that best fits their needs and preferences. Consequently, we argue that, regardless of the users’ background knowledge, empowering users to take control of the recommendation should be an integral feature in any ERS, if transparency is a desirable
property for the ERS. We refer to this type of transparency as transparency through controllability.

Furthermore, our analysis shows that user control moderately correlates with trust with a statistical significance. This confirms that providing more control over RS is an important strategy to secure trust \cite{harambam,jannach2017user,jannach2019explanations} and that interactive features that allow users to control the RS increase their trust in the system \cite{schaffer2015hypothetical, bruns2015should, loepp2014choice}.

Our study further shows that user control moderately correlates with user satisfaction with a statistical significance. This suggests that user control over the RS positively influenced their satisfaction, indicating an enhanced user experience. Our results confirm the findings by \citet{Pu2011resque}, who noted that user control weighs heavily on the overall user experience with the RS. Furthermore, several studies from the recommendation domain evidenced the positive effects of user control on satisfaction \cite{tsai2017providing, jin2016go, linkedvis2013, smallworlds, talkexplorer} and user experience \cite{knijnenburg2012inspectability, Donovan, 2012tasteweights, setfusion2014}.
In summary, our findings clearly demonstrate the positive impact of user control in ERS, showing that providing user control over the ERS leads to increased perceived transparency, trust, and satisfaction.
\subsection{Interaction Between Different Goals} \label{goals}
Regarding RQ3, the results from the correlation analysis (see Figure \ref{correlation}) reveal that all the three goals, transparency, trust, and satisfaction appear to move together, however, the correlation is not strong in all cases. We found in our study that transparency moderately correlates with satisfaction with a statistical significance. Moreover, trust strongly correlates with satisfaction with a statistical significance. We conclude that satisfaction is most correlated metric with all goals. This aligns with previous research suggesting that overall user satisfaction with RS is strongly related to transparency and trust. For instance, in the explainable RS domain, \citet{balog2020measuring}, and \citet{guesmi2023interactive, guesmi2023justification} found that satisfaction is positively correlated with transparency and trust. In the same context, \citet{gedikli2014should} reported
results from experiments with different explanations clearly showing that transparency has a significant positive effect on user satisfaction. Moreover, a lower-transparency RS is known to negatively affect user satisfaction with the RS \cite{tintarev2015explaining}.
Overall, our results provide evidence that, similar to  findings from the explainable RS domain, transparency and trust have significant positive effects on user satisfaction in the interactive RS domain as well.

Additionally, we found in our study that transparency and trust stand out as less correlated with each other. This suggests that transparency and trust, if they are desirable properties for an IntRS, should be evaluated as distinct, even if they may interact. One reason we consider responsible for the relative low correlation between transparency and trust is that offering interaction mechanisms at different levels provides a significant degree of user control which is, as discussed in Section \ref{impact}, highly influential to transparency, but also introduces the risk of overwhelming users \cite{jannach2019explanations}. As pointed out by \citet{jin2018effects}, providing additional controls can also increase cognitive load, and different users have different needs for control. This suggests that providing too much control does not necessarily lead to greater trust, as it can increase the cognitive load on users. This further highlights that there is a trade-off between the amount of user control and the level of trust users develop when interacting with the RS. Further research is needed to find an optimal level of user control that will generate the highest
level of users’ trust in the RS. 

Another reason that can explain the relatively low correlation between transparency and trust is that while transparency can be achieved through controllability, interacting with the input, process, or output of the RS cannot always assure that users understand the underlying
rationale of the RS, especially when the recommendation mechanism is too complicated to non-expert users \cite{tsai2021effects}, which can negatively impact the perceived trust of the RS.  
Some considerable transparency could
be achieved through explanation which is also recognized as an important factor that fosters user trust in RS, as it can improve users' understanding of how the system works \cite{tintarev2015explaining}.
We hypothesize that a different result may be observed if explanations were provided in the ERS together with interactions. It is therefore important to explore in the future work  the effects of different types of transparency (i.e., transparency through controllability vs. transparency through explanation) on users’ trust in RSs \cite{siepmann2023trust}.

A further possible reason for the relative low correlation between transparency and trust is that, as found in \cite{Pu2011resque}, trust depends primarily on the RS’s ability to formulate good
recommendations (i.e., perceived usefulness) and provide useful explanations (i.e., transparency through explanation), rather than on control (i.e., transparency through controllability). 
\section{Conclusion and Future Work} \label{conclusion}
In this paper, we aimed to shed light on an aspect that remains
under-researched in the literature on educational recommender systems (ERSs),
namely the effects of providing user control on users’ perceptions of ERSs. To
this end, we systematically designed and evaluated user control in the ERS module of the MOOC platform CourseMapper. Specifically, we introduced user control with the input (i.e., user profile), process (i.e., recommendation algorithm), and output (i.e., recommendations) of the ERS. We conducted an online user study (N=30) to explore the impact of user control
on users’ perceptions of the ERS in terms of several important user-centric aspects. Moreover, we examined the impact of user control on transparency, trust, and satisfaction, and explored the interactions among these recommendation goals. Our findings indicate that users responded positively to the control mechanisms provided in the ERS. Our results further reveal that satisfaction is the most correlated metric with all goals and that transparency and trust stand out as less correlated with each other. This suggests
that it may be necessary to separately consider transparency and trust in the evaluation of interactive ERSs.
As a future work, we plan to conduct a more comprehensive user study, incorporating qualitative analysis to gain deeper insights into the relationships between user control, transparency, trust, and satisfaction. An interesting
direction in future work would be to also investigate the effects of different levels of user control (i.e., input, process, output) on the users' perceptions of and interactions with the ERS.
\bibliography{sample-1col}

\begin{thebibliography}{52}
\expandafter\ifx\csname natexlab\endcsname\relax\def\natexlab#1{#1}\fi
\providecommand{\url}[1]{\texttt{#1}}
\providecommand{\href}[2]{#2}
\providecommand{\path}[1]{#1}
\providecommand{\DOIprefix}{doi:}
\providecommand{\ArXivprefix}{arXiv:}
\providecommand{\URLprefix}{URL: }
\providecommand{\Pubmedprefix}{pmid:}
\providecommand{\doi}[1]{\href{http://dx.doi.org/#1}{\path{#1}}}
\providecommand{\Pubmed}[1]{\href{pmid:#1}{\path{#1}}}
\providecommand{\bibinfo}[2]{#2}
\ifx\xfnm\relax \def\xfnm[#1]{\unskip,\space#1}\fi
\bibitem[{Manouselis et~al.(2011)Manouselis, Drachsler, Vuorikari, Hummel, and
  Koper}]{manouselis2011recommender}
\bibinfo{author}{N.~Manouselis}, \bibinfo{author}{H.~Drachsler},
  \bibinfo{author}{R.~Vuorikari}, \bibinfo{author}{H.~Hummel},
  \bibinfo{author}{R.~Koper},
\newblock \bibinfo{title}{Recommender systems in technology enhanced learning},
\newblock \bibinfo{journal}{Recommender systems handbook}
  (\bibinfo{year}{2011}) \bibinfo{pages}{387--415}.
\bibitem[{Khanal et~al.(2020)Khanal, Prasad, Alsadoon, and
  Maag}]{khanal2020systematic}
\bibinfo{author}{S.~S. Khanal}, \bibinfo{author}{P.~Prasad},
  \bibinfo{author}{A.~Alsadoon}, \bibinfo{author}{A.~Maag},
\newblock \bibinfo{title}{A systematic review: machine learning based
  recommendation systems for e-learning},
\newblock \bibinfo{journal}{Education and Information Technologies}
  (\bibinfo{year}{2020}).
\bibitem[{Valtolina et~al.(2024)Valtolina, Matamoros, and
  Epifania}]{valtolina2024design}
\bibinfo{author}{S.~Valtolina}, \bibinfo{author}{R.~A. Matamoros},
  \bibinfo{author}{F.~Epifania},
\newblock \bibinfo{title}{Design of a conversational recommender system in
  education},
\newblock \bibinfo{journal}{User modeling and user-adapted interaction}
  (\bibinfo{year}{2024}) \bibinfo{pages}{1--29}.
\bibitem[{Chau et~al.(2018)Chau, Barria-Pineda, and
  Brusilovsky}]{chau2018learning}
\bibinfo{author}{H.~Chau}, \bibinfo{author}{J.~Barria-Pineda},
  \bibinfo{author}{P.~Brusilovsky},
\newblock \bibinfo{title}{Learning content recommender system for instructors
  of programming courses},
\newblock in: \bibinfo{booktitle}{Artificial Intelligence in Education: 19th
  International Conference, AIED 2018, London, UK, June 27--30, 2018,
  Proceedings, Part II 19}, \bibinfo{organization}{Springer},
  \bibinfo{year}{2018}, pp. \bibinfo{pages}{47--51}.
\bibitem[{Bousbahi and Chorfi(2015)}]{BOUSBAHI20151813}
\bibinfo{author}{F.~Bousbahi}, \bibinfo{author}{H.~Chorfi},
\newblock \bibinfo{title}{Mooc-rec: A case based recommender system for moocs},
\newblock \bibinfo{journal}{Procedia - Social and Behavioral Sciences}
  \bibinfo{volume}{195} (\bibinfo{year}{2015}) \bibinfo{pages}{1813--1822}.
  \bibinfo{note}{World Conference on Technology, Innovation and
  Entrepreneurship}.
\bibitem[{Santos et~al.(2016)Santos, Saneiro, Boticario, and
  Rodriguez-Sanchez}]{santos2016toward}
\bibinfo{author}{O.~C. Santos}, \bibinfo{author}{M.~Saneiro},
  \bibinfo{author}{J.~G. Boticario}, \bibinfo{author}{M.~C. Rodriguez-Sanchez},
\newblock \bibinfo{title}{Toward interactive context-aware affective
  educational recommendations in computer-assisted language learning},
\newblock \bibinfo{journal}{New Review of Hypermedia and Multimedia}
  \bibinfo{volume}{22} (\bibinfo{year}{2016}) \bibinfo{pages}{27--57}.
\bibitem[{Jin et~al.(2018)Jin, Tintarev, and Verbert}]{jin2018effects}
\bibinfo{author}{Y.~Jin}, \bibinfo{author}{N.~Tintarev},
  \bibinfo{author}{K.~Verbert},
\newblock \bibinfo{title}{Effects of personal characteristics on music
  recommender systems with different levels of controllability},
\newblock in: \bibinfo{booktitle}{Proceedings of the 12th ACM Conference on
  Recommender Systems}, \bibinfo{year}{2018}, pp. \bibinfo{pages}{13--21}.
\bibitem[{He et~al.(2016)He, Parra, and Verbert}]{he2016interactive}
\bibinfo{author}{C.~He}, \bibinfo{author}{D.~Parra},
  \bibinfo{author}{K.~Verbert},
\newblock \bibinfo{title}{Interactive recommender systems: A survey of the
  state of the art and future research challenges and opportunities},
\newblock \bibinfo{journal}{Expert Systems with Applications}
  \bibinfo{volume}{56} (\bibinfo{year}{2016}) \bibinfo{pages}{9--27}.
\bibitem[{Jugovac and Jannach(2017)}]{jugovac2017interacting}
\bibinfo{author}{M.~Jugovac}, \bibinfo{author}{D.~Jannach},
\newblock \bibinfo{title}{Interacting with recommenders—overview and research
  directions},
\newblock \bibinfo{journal}{ACM Transactions on Interactive Intelligent Systems
  (TiiS)} \bibinfo{volume}{7} (\bibinfo{year}{2017}) \bibinfo{pages}{1--46}.
\bibitem[{Jannach et~al.(2017)Jannach, Naveed, and Jugovac}]{jannach2017user}
\bibinfo{author}{D.~Jannach}, \bibinfo{author}{S.~Naveed},
  \bibinfo{author}{M.~Jugovac},
\newblock \bibinfo{title}{User control in recommender systems: Overview and
  interaction challenges},
\newblock in: \bibinfo{booktitle}{E-Commerce and Web Technologies: 17th
  International Conference, EC-Web 2016, Porto, Portugal, September 5-8, 2016,
  Revised Selected Papers 17}, \bibinfo{organization}{Springer},
  \bibinfo{year}{2017}, pp. \bibinfo{pages}{21--33}.
\bibitem[{Harambam et~al.(2019)Harambam, Bountouridis, Makhortykh, and van
  Hoboken}]{harambam}
\bibinfo{author}{J.~Harambam}, \bibinfo{author}{D.~Bountouridis},
  \bibinfo{author}{M.~Makhortykh}, \bibinfo{author}{J.~van Hoboken},
\newblock \bibinfo{title}{Designing for the better by taking users into
  account: a qualitative evaluation of user control mechanisms in (news)
  recommender systems},
\newblock in: \bibinfo{booktitle}{Proceedings of 13th ACM Conference on
  Recommender Systems}, RecSys '19, \bibinfo{year}{2019}.
\bibitem[{Knijnenburg et~al.(2012)Knijnenburg, Bostandjiev, O'Donovan, and
  Kobsa}]{knijnenburg2012inspectability}
\bibinfo{author}{B.~P. Knijnenburg}, \bibinfo{author}{S.~Bostandjiev},
  \bibinfo{author}{J.~O'Donovan}, \bibinfo{author}{A.~Kobsa},
\newblock \bibinfo{title}{Inspectability and control in social recommenders},
\newblock in: \bibinfo{booktitle}{Proceedings of the sixth ACM conference on
  Recommender systems}, \bibinfo{year}{2012}.
\bibitem[{Barria-Pineda and Brusilovsky(2019)}]{barria2019explaining}
\bibinfo{author}{J.~Barria-Pineda}, \bibinfo{author}{P.~Brusilovsky},
\newblock \bibinfo{title}{Explaining educational recommendations through a
  concept-level knowledge visualization},
\newblock in: \bibinfo{booktitle}{Companion Proceedings of the 24th
  International Conference on Intelligent User Interfaces},
  \bibinfo{year}{2019}, pp. \bibinfo{pages}{103--104}.
\bibitem[{Schaffer et~al.(2015)Schaffer, Hollerer, and
  O'Donovan}]{schaffer2015hypothetical}
\bibinfo{author}{J.~Schaffer}, \bibinfo{author}{T.~Hollerer},
  \bibinfo{author}{J.~O'Donovan},
\newblock \bibinfo{title}{Hypothetical recommendation: A study of interactive
  profile manipulation behavior for recommender systems},
\newblock in: \bibinfo{booktitle}{28th international flairs conference},
  \bibinfo{year}{2015}.
\bibitem[{O'Donovan et~al.(2008)O'Donovan, Smyth, Gretarsson, Bostandjiev, and
  H\"{o}llerer}]{Donovan}
\bibinfo{author}{J.~O'Donovan}, \bibinfo{author}{B.~Smyth},
  \bibinfo{author}{B.~Gretarsson}, \bibinfo{author}{S.~Bostandjiev},
  \bibinfo{author}{T.~H\"{o}llerer},
\newblock \bibinfo{title}{Peerchooser: visual interactive recommendation},
\newblock in: \bibinfo{booktitle}{Proceedings of the SIGCHI Conference on Human
  Factors in Computing Systems}, CHI '08, \bibinfo{publisher}{Association for
  Computing Machinery}, \bibinfo{address}{NY, USA}, \bibinfo{year}{2008}, p.
  \bibinfo{pages}{1085–1088}.
\bibitem[{Bostandjiev et~al.(2012)Bostandjiev, O'Donovan, and
  H\"{o}llerer}]{2012tasteweights}
\bibinfo{author}{S.~Bostandjiev}, \bibinfo{author}{J.~O'Donovan},
  \bibinfo{author}{T.~H\"{o}llerer},
\newblock \bibinfo{title}{Tasteweights: a visual interactive hybrid recommender
  system},
\newblock in: \bibinfo{booktitle}{Proceedings of the 6th ACM Conference on
  Recommender Systems}, \bibinfo{publisher}{Association for Computing
  Machinery}, \bibinfo{address}{New York, USA}, \bibinfo{year}{2012}.
\bibitem[{Harper et~al.(2015)Harper, Xu, Kaur, Condiff, Chang, and
  Terveen}]{maxwell2015}
\bibinfo{author}{F.~M. Harper}, \bibinfo{author}{F.~Xu},
  \bibinfo{author}{H.~Kaur}, \bibinfo{author}{K.~Condiff},
  \bibinfo{author}{S.~Chang}, \bibinfo{author}{L.~Terveen},
\newblock \bibinfo{title}{Putting users in control of their recommendations},
\newblock in: \bibinfo{booktitle}{Proceedings of the 9th ACM Conference on
  Recommender Systems}, RecSys '15, \bibinfo{publisher}{Association for
  Computing Machinery}, \bibinfo{address}{New York, NY, USA},
  \bibinfo{year}{2015}, p. \bibinfo{pages}{3–10}.
\bibitem[{Gretarsson et~al.(2010)Gretarsson, O'Donovan, Bostandjiev, Hall, and
  Höllerer}]{smallworlds}
\bibinfo{author}{B.~Gretarsson}, \bibinfo{author}{J.~O'Donovan},
  \bibinfo{author}{S.~Bostandjiev}, \bibinfo{author}{C.~Hall},
  \bibinfo{author}{T.~Höllerer},
\newblock \bibinfo{title}{Smallworlds: Visualizing social recommendations},
\newblock \bibinfo{journal}{Computer Graphics Forum} \bibinfo{volume}{29}
  (\bibinfo{year}{2010}) \bibinfo{pages}{833--842}.
\bibitem[{Tsai and Brusilovsky(2017)}]{tsai2017providing}
\bibinfo{author}{C.-H. Tsai}, \bibinfo{author}{P.~Brusilovsky},
\newblock \bibinfo{title}{Providing control and transparency in a social
  recommender system for academic conferences},
\newblock in: \bibinfo{booktitle}{Proceedings of the 25th conference on user
  modeling, adaptation and personalization}, \bibinfo{year}{2017}, pp.
  \bibinfo{pages}{313--317}.
\bibitem[{Kangasr\"{a}\"{a}si\"{o} et~al.(2015)Kangasr\"{a}\"{a}si\"{o},
  Glowacka, and Kaski}]{Kangasrasio2015}
\bibinfo{author}{A.~Kangasr\"{a}\"{a}si\"{o}}, \bibinfo{author}{D.~Glowacka},
  \bibinfo{author}{S.~Kaski},
\newblock \bibinfo{title}{Improving controllability and predictability of
  interactive recommendation interfaces for exploratory search},
\newblock in: \bibinfo{booktitle}{Proceedings of the 20th International
  Conference on Intelligent User Interfaces}, IUI '15, \bibinfo{year}{2015}.
\bibitem[{Bruns et~al.(2015)Bruns, Valdez, Greven, Ziefle, and
  Schroeder}]{bruns2015should}
\bibinfo{author}{S.~Bruns}, \bibinfo{author}{A.~C. Valdez},
  \bibinfo{author}{C.~Greven}, \bibinfo{author}{M.~Ziefle},
  \bibinfo{author}{U.~Schroeder},
\newblock \bibinfo{title}{What should i read next? a personalized visual
  publication recommender system},
\newblock in: \bibinfo{booktitle}{International Conference on Human Interface
  and the Management of Information}, \bibinfo{organization}{Springer},
  \bibinfo{year}{2015}, pp. \bibinfo{pages}{89--100}.
\bibitem[{Zhao et~al.(2010)Zhao, Zhou, Yuan, Zhang, and Zheng}]{zhao2010}
\bibinfo{author}{S.~Zhao}, \bibinfo{author}{M.~X. Zhou},
  \bibinfo{author}{Q.~Yuan}, \bibinfo{author}{X.~Zhang},
  \bibinfo{author}{R.~Zheng},
\newblock \bibinfo{title}{Who is talking about what: social map-based
  recommendation for content-centric social websites},
\newblock in: \bibinfo{booktitle}{Proceedings of the 4th ACM Conference on
  Recommender Systems}, \bibinfo{address}{New York, USA}, \bibinfo{year}{2010}.
\bibitem[{Tintarev et~al.(2015)Tintarev, Kang, H{\"o}llerer, and
  O'Donovan}]{tintarev2015inspection}
\bibinfo{author}{N.~Tintarev}, \bibinfo{author}{B.~Kang},
  \bibinfo{author}{T.~H{\"o}llerer}, \bibinfo{author}{J.~O'Donovan},
\newblock \bibinfo{title}{Inspection mechanisms for community-based content
  discovery in microblogs.},
\newblock in: \bibinfo{booktitle}{IntRS@ RecSys}, \bibinfo{year}{2015}, pp.
  \bibinfo{pages}{21--28}.
\bibitem[{Chen and Pu(2012)}]{chen2012cofeel}
\bibinfo{author}{Y.~Chen}, \bibinfo{author}{P.~Pu},
\newblock \bibinfo{title}{Cofeel: Using emotions for social interaction in
  group recommender systems}  (\bibinfo{year}{2012}).
\bibitem[{Tsai and Brusilovsky(2021)}]{tsai2021effects}
\bibinfo{author}{C.-H. Tsai}, \bibinfo{author}{P.~Brusilovsky},
\newblock \bibinfo{title}{The effects of controllability and explainability in
  a social recommender system},
\newblock \bibinfo{journal}{User Modeling and User-Adapted Interaction}
  \bibinfo{volume}{31} (\bibinfo{year}{2021}) \bibinfo{pages}{591--627}.
\bibitem[{Loepp et~al.(2014)Loepp, Hussein, and Ziegler}]{loepp2014choice}
\bibinfo{author}{B.~Loepp}, \bibinfo{author}{T.~Hussein},
  \bibinfo{author}{J.~Ziegler},
\newblock \bibinfo{title}{Choice-based preference elicitation for collaborative
  filtering recommender systems},
\newblock in: \bibinfo{booktitle}{Proceedings of the SIGCHI Conference on Human
  Factors in Computing Systems}, \bibinfo{year}{2014}, pp.
  \bibinfo{pages}{3085--3094}.
\bibitem[{Parra et~al.(2014)Parra, Brusilovsky, and Trattner}]{setfusion2014}
\bibinfo{author}{D.~Parra}, \bibinfo{author}{P.~Brusilovsky},
  \bibinfo{author}{C.~Trattner},
\newblock \bibinfo{title}{See what you want to see: visual user-driven approach
  for hybrid recommendation},
\newblock in: \bibinfo{booktitle}{Proceedings of the 19th International
  Conference on Intelligent User Interfaces}, IUI '14,
  \bibinfo{publisher}{Association for Computing Machinery},
  \bibinfo{address}{New York, USA}, \bibinfo{year}{2014}, p.
  \bibinfo{pages}{235–240}.
\bibitem[{Jin et~al.(2016)Jin, Seipp, Duval, and Verbert}]{jin2016go}
\bibinfo{author}{Y.~Jin}, \bibinfo{author}{K.~Seipp},
  \bibinfo{author}{E.~Duval}, \bibinfo{author}{K.~Verbert},
\newblock \bibinfo{title}{Go with the flow: effects of transparency and user
  control on targeted advertising using flow charts},
\newblock in: \bibinfo{booktitle}{Proceedings of the international working
  conference on advanced visual interfaces}, \bibinfo{year}{2016}, pp.
  \bibinfo{pages}{68--75}.
\bibitem[{Bostandjiev et~al.(2013)Bostandjiev, O'Donovan, and
  Höllerer}]{linkedvis2013}
\bibinfo{author}{S.~Bostandjiev}, \bibinfo{author}{J.~O'Donovan},
  \bibinfo{author}{T.~Höllerer},
\newblock \bibinfo{title}{Linkedvis: Exploring social and semantic career
  recommendations},
\newblock \bibinfo{year}{2013}, pp. \bibinfo{pages}{107--116}.
  \DOIprefix\doi{10.1145/2449396.2449412}.
\bibitem[{Verbert et~al.(2013)Verbert, Parra, Brusilovsky, and
  Duval}]{talkexplorer}
\bibinfo{author}{K.~Verbert}, \bibinfo{author}{D.~Parra},
  \bibinfo{author}{P.~Brusilovsky}, \bibinfo{author}{E.~Duval},
\newblock \bibinfo{title}{Visualizing recommendations to support exploration,
  transparency and controllability},
\newblock in: \bibinfo{booktitle}{Proceedings of the 2013 International
  Conference on Intelligent User Interfaces}, IUI '13,
  \bibinfo{publisher}{ACM}, \bibinfo{address}{New York, NY, USA},
  \bibinfo{year}{2013}.
\bibitem[{Tintarev and Masthoff(2015)}]{tintarev2015explaining}
\bibinfo{author}{N.~Tintarev}, \bibinfo{author}{J.~Masthoff},
\newblock \bibinfo{title}{Explaining recommendations: Design and evaluation},
\newblock in: \bibinfo{booktitle}{Recommender systems handbook},
  \bibinfo{publisher}{Springer}, \bibinfo{year}{2015}, pp.
  \bibinfo{pages}{353--382}.
\bibitem[{Ain et~al.(2023)Ain, Chatti, Joarder, Nassif, Wobiwo~Teda, Guesmi,
  and Alatrash}]{Ain2022}
\bibinfo{author}{Q.~U. Ain}, \bibinfo{author}{M.~A. Chatti},
  \bibinfo{author}{S.~Joarder}, \bibinfo{author}{I.~Nassif},
  \bibinfo{author}{B.~S. Wobiwo~Teda}, \bibinfo{author}{M.~Guesmi},
  \bibinfo{author}{R.~Alatrash},
\newblock \bibinfo{title}{Learning channels to support interaction and
  collaboration in coursemapper},
\newblock in: \bibinfo{booktitle}{Proceedings of the 14th International
  Conference on Education Technology and Computers}, ICETC '22,
  \bibinfo{year}{2023}.
\bibitem[{Ain et~al.(2024)Ain, Chatti, Meteng~Kamdem, Alatrash, Joarder, and
  Siepmann}]{ain2024learner}
\bibinfo{author}{Q.~U. Ain}, \bibinfo{author}{M.~A. Chatti},
  \bibinfo{author}{P.~A. Meteng~Kamdem}, \bibinfo{author}{R.~Alatrash},
  \bibinfo{author}{S.~Joarder}, \bibinfo{author}{C.~Siepmann},
\newblock \bibinfo{title}{Learner modeling and recommendation of learning
  resources using personal knowledge graphs},
\newblock in: \bibinfo{booktitle}{Proceedings of the 14th Learning Analytics
  and Knowledge Conference}, \bibinfo{year}{2024}, pp.
  \bibinfo{pages}{273--283}.
\bibitem[{Fotopoulou et~al.(2020)Fotopoulou, Zafeiropoulos, Feidakis, Metafas,
  and Papavassiliou}]{fotopoulou2020interactive}
\bibinfo{author}{E.~Fotopoulou}, \bibinfo{author}{A.~Zafeiropoulos},
  \bibinfo{author}{M.~Feidakis}, \bibinfo{author}{D.~Metafas},
  \bibinfo{author}{S.~Papavassiliou},
\newblock \bibinfo{title}{An interactive recommender system based on
  reinforcement learning for improving emotional competences in educational
  groups},
\newblock in: \bibinfo{booktitle}{International Conference on Intelligent
  Tutoring Systems}, \bibinfo{organization}{Springer}, \bibinfo{year}{2020}.
\bibitem[{Bustos~L{\'o}pez et~al.(2020)Bustos~L{\'o}pez, Alor-Hern{\'a}ndez,
  S{\'a}nchez-Cervantes, Paredes-Valverde, and
  Salas-Z{\'a}rate}]{bustos2020edurecomsys}
\bibinfo{author}{M.~Bustos~L{\'o}pez}, \bibinfo{author}{G.~Alor-Hern{\'a}ndez},
  \bibinfo{author}{J.~L. S{\'a}nchez-Cervantes}, \bibinfo{author}{M.~A.
  Paredes-Valverde}, \bibinfo{author}{M.~d.~P. Salas-Z{\'a}rate},
\newblock \bibinfo{title}{Edurecomsys: an educational resource recommender
  system based on collaborative filtering and emotion detection},
\newblock \bibinfo{journal}{Interacting with Computers} \bibinfo{volume}{32}
  (\bibinfo{year}{2020}) \bibinfo{pages}{407--432}.
\bibitem[{da~Silva et~al.(2023)da~Silva, Slodkowski, da~Silva, and
  Cazella}]{da2023systematic}
\bibinfo{author}{F.~L. da~Silva}, \bibinfo{author}{B.~K. Slodkowski},
  \bibinfo{author}{K.~K.~A. da~Silva}, \bibinfo{author}{S.~C. Cazella},
\newblock \bibinfo{title}{A systematic literature review on educational
  recommender systems for teaching and learning: research trends, limitations
  and opportunities},
\newblock \bibinfo{journal}{Education and Information Technologies}
  \bibinfo{volume}{28} (\bibinfo{year}{2023}) \bibinfo{pages}{3289--3328}.
\bibitem[{Zapata et~al.(2015)Zapata, Men{\'e}ndez, Prieto, and
  Romero}]{zapata2015evaluation}
\bibinfo{author}{A.~Zapata}, \bibinfo{author}{V.~H. Men{\'e}ndez},
  \bibinfo{author}{M.~E. Prieto}, \bibinfo{author}{C.~Romero},
\newblock \bibinfo{title}{Evaluation and selection of group recommendation
  strategies for collaborative searching of learning objects},
\newblock \bibinfo{journal}{International Journal of Human-Computer Studies}
  \bibinfo{volume}{76} (\bibinfo{year}{2015}) \bibinfo{pages}{22--39}.
\bibitem[{Abdi et~al.(2020)Abdi, Khosravi, Sadiq, and
  Gasevic}]{abdi2020complementing}
\bibinfo{author}{S.~Abdi}, \bibinfo{author}{H.~Khosravi},
  \bibinfo{author}{S.~Sadiq}, \bibinfo{author}{D.~Gasevic},
\newblock \bibinfo{title}{Complementing educational recommender systems with
  open learner models},
\newblock in: \bibinfo{booktitle}{Proceedings of the tenth international
  conference on learning analytics \& knowledge}, \bibinfo{year}{2020}, pp.
  \bibinfo{pages}{360--365}.
\bibitem[{Vlachos and Svonava(2012)}]{Svonava2012}
\bibinfo{author}{M.~Vlachos}, \bibinfo{author}{D.~Svonava},
\newblock \bibinfo{title}{Graph embeddings for movie visualization and
  recommendation},
\newblock in: \bibinfo{booktitle}{First International Workshop on
  Recommendation Technologies for Lifestyle Change}, \bibinfo{year}{2012}.
\bibitem[{Schafer et~al.(2002)Schafer, Konstan, and Riedl}]{schafer2002}
\bibinfo{author}{J.~B. Schafer}, \bibinfo{author}{J.~A. Konstan},
  \bibinfo{author}{J.~Riedl},
\newblock \bibinfo{title}{Meta-recommendation systems: user-controlled
  integration of diverse recommendations},
\newblock in: \bibinfo{booktitle}{Proceedings of the Eleventh International
  Conference on Information and Knowledge Management},
  \bibinfo{publisher}{Association for Computing Machinery},
  \bibinfo{address}{NY, USA}, \bibinfo{year}{2002}.
\bibitem[{Saito and Itoh(2011)}]{saito2011}
\bibinfo{author}{Y.~Saito}, \bibinfo{author}{T.~Itoh},
\newblock \bibinfo{title}{Musicube: a visual music recommendation system
  featuring interactive evolutionary computing},
\newblock in: \bibinfo{booktitle}{Proceedings of the 2011 Visual Information
  Communication - International Symposium}, VINCI '11,
  \bibinfo{publisher}{Association for Computing Machinery},
  \bibinfo{address}{New York, USA}, \bibinfo{year}{2011}.
\bibitem[{Wong et~al.(2011)Wong, Faridani, Bitton, Hartmann, and
  Goldberg}]{wong2011diversity}
\bibinfo{author}{D.~Wong}, \bibinfo{author}{S.~Faridani},
  \bibinfo{author}{E.~Bitton}, \bibinfo{author}{B.~Hartmann},
  \bibinfo{author}{K.~Goldberg},
\newblock \bibinfo{title}{The diversity donut: enabling participant control
  over the diversity of recommended responses},
\newblock in: \bibinfo{booktitle}{CHI'11 Extended Abstracts on Human Factors in
  Computing Systems}, \bibinfo{year}{2011}, pp. \bibinfo{pages}{1471--1476}.
\bibitem[{Ekstrand et~al.(2015)Ekstrand, Kluver, Harper, and
  Konstan}]{Ekstrand2015LettingUC}
\bibinfo{author}{M.~D. Ekstrand}, \bibinfo{author}{D.~Kluver},
  \bibinfo{author}{F.~M. Harper}, \bibinfo{author}{J.~A. Konstan},
\newblock \bibinfo{title}{Letting users choose recommender algorithms: An
  experimental study},
\newblock \bibinfo{journal}{Proceedings of the 9th ACM Conference on
  Recommender Systems}  (\bibinfo{year}{2015}).
\bibitem[{Ooge et~al.(2023)Ooge, Dereu, and Verbert}]{ooge23steering}
\bibinfo{author}{J.~Ooge}, \bibinfo{author}{L.~Dereu},
  \bibinfo{author}{K.~Verbert},
\newblock \bibinfo{title}{Steering recommendations and visualising its impact:
  Effects on adolescents’ trust in e-learning platforms},
\newblock in: \bibinfo{booktitle}{Proceedings of the 28th International
  Conference on Intelligent User Interfaces}, IUI '23, \bibinfo{year}{2023}, p.
  \bibinfo{pages}{156–170}.
\bibitem[{Jannach et~al.(2019)Jannach, Jugovac, and
  Nunes}]{jannach2019explanations}
\bibinfo{author}{D.~Jannach}, \bibinfo{author}{M.~Jugovac},
  \bibinfo{author}{I.~Nunes},
\newblock \bibinfo{title}{Explanations and user control in recommender
  systems},
\newblock in: \bibinfo{booktitle}{Proceedings of the 23rd International
  Workshop on Personalization and Recommendation on the Web and Beyond},
  \bibinfo{year}{2019}, pp. \bibinfo{pages}{31--31}.
\bibitem[{Pu et~al.(2011)Pu, Chen, and Hu}]{Pu2011resque}
\bibinfo{author}{P.~Pu}, \bibinfo{author}{L.~Chen}, \bibinfo{author}{R.~Hu},
\newblock \bibinfo{title}{A user-centric evaluation framework for recommender
  systems},
\newblock in: \bibinfo{booktitle}{ACM Conference on Recommender Systems},
  \bibinfo{year}{2011}.
\bibitem[{Tintarev and Masthoff(2007)}]{tintarev2007}
\bibinfo{author}{N.~Tintarev}, \bibinfo{author}{J.~Masthoff},
\newblock \bibinfo{title}{A survey of explanations in recommender systems},
\newblock in: \bibinfo{booktitle}{2007 IEEE 23rd International Conference on
  Data Engineering Workshop}, \bibinfo{year}{2007}, pp.
  \bibinfo{pages}{801--810}.
\bibitem[{Balog and Radlinski(2020)}]{balog2020measuring}
\bibinfo{author}{K.~Balog}, \bibinfo{author}{F.~Radlinski},
\newblock \bibinfo{title}{Measuring recommendation explanation quality: The
  conflicting goals of explanations},
\newblock in: \bibinfo{booktitle}{Proceedings of the 43rd international ACM
  SIGIR conference on research and development in information retrieval},
  \bibinfo{year}{2020}, pp. \bibinfo{pages}{329--338}.
\bibitem[{Guesmi et~al.(2023{\natexlab{a}})Guesmi, Chatti, Joarder, Ain,
  Alatrash, Siepmann, and Vahidi}]{guesmi2023interactive}
\bibinfo{author}{M.~Guesmi}, \bibinfo{author}{M.~A. Chatti},
  \bibinfo{author}{S.~Joarder}, \bibinfo{author}{Q.~U. Ain},
  \bibinfo{author}{R.~Alatrash}, \bibinfo{author}{C.~Siepmann},
  \bibinfo{author}{T.~Vahidi},
\newblock \bibinfo{title}{Interactive explanation with varying level of details
  in an explainable scientific literature recommender system},
\newblock \bibinfo{journal}{International Journal of Human--Computer
  Interaction}  (\bibinfo{year}{2023}{\natexlab{a}}) \bibinfo{pages}{1--22}.
\bibitem[{Guesmi et~al.(2023{\natexlab{b}})Guesmi, Chatti, Joarder, Ain,
  Siepmann, Ghanbarzadeh, and Alatrash}]{guesmi2023justification}
\bibinfo{author}{M.~Guesmi}, \bibinfo{author}{M.~A. Chatti},
  \bibinfo{author}{S.~Joarder}, \bibinfo{author}{Q.~U. Ain},
  \bibinfo{author}{C.~Siepmann}, \bibinfo{author}{H.~Ghanbarzadeh},
  \bibinfo{author}{R.~Alatrash},
\newblock \bibinfo{title}{Justification vs. transparency: Why and how visual
  explanations in a scientific literature recommender system},
\newblock \bibinfo{journal}{Information} \bibinfo{volume}{14}
  (\bibinfo{year}{2023}{\natexlab{b}}) \bibinfo{pages}{401}.
\bibitem[{Gedikli et~al.(2014)Gedikli, Jannach, and Ge}]{gedikli2014should}
\bibinfo{author}{F.~Gedikli}, \bibinfo{author}{D.~Jannach},
  \bibinfo{author}{M.~Ge},
\newblock \bibinfo{title}{How should i explain? a comparison of different
  explanation types for recommender systems},
\newblock \bibinfo{journal}{International Journal of Human-Computer Studies}
  \bibinfo{volume}{72} (\bibinfo{year}{2014}) \bibinfo{pages}{367--382}.
\bibitem[{Siepmann and Chatti(2023)}]{siepmann2023trust}
\bibinfo{author}{C.~Siepmann}, \bibinfo{author}{M.~A. Chatti},
\newblock \bibinfo{title}{Trust and transparency in recommender systems},
\newblock \bibinfo{journal}{arXiv preprint arXiv:2304.08094}
  (\bibinfo{year}{2023}).

\end{thebibliography}

\end{document}